\documentclass[12pt]{article}
\usepackage{amsmath,amssymb,amsfonts,graphicx,verbatim,cite,tensor}
\usepackage{wrapfig}

\pdfoutput=1

\usepackage[usenames,dvipsnames]{xcolor}

\usepackage[pdftex, bookmarks={false},
pdfauthor={Brian Swingle}, pdftitle={notes}]{hyperref}
\hypersetup{colorlinks=true, linkcolor=BrickRed, citecolor=Violet, filecolor=OliveGreen, urlcolor=RoyalBlue, filebordercolor={.8 .8 1}, urlbordercolor={.8 .8 0}}

\definecolor{palatinatepurple}{rgb}{0.41, 0.16, 0.38}

\definecolor{uglybrown}{rgb}{0.8,  0.7,  0.5}

\renewcommand{\title}[1]{\vbox{\center\LARGE{#1}}\vspace{5mm}}
\renewcommand{\author}[1]{\vbox{\center#1}\vspace{5mm}}
\newcommand{\address}[1]{\vbox{\center\em#1}}

\numberwithin{equation}{section}

\newcommand{\beq}{\begin{equation}}
\newcommand{\eeq}{\end{equation}}
\def\bea{\begin{eqnarray}}
\def\eea{\end{eqnarray}}

\def\({\left(}
\def\){\right)}

\def\CC{{\cal C}}

\def\CL{{\cal L}}

\def\CN{{\cal N}}

\def\CQ{{\cal Q}}
\def\CR{{\cal R}}
\def\CS{{\cal S}}

\def\CV{{\cal V}}

\usepackage{tikz}
\usetikzlibrary{calc,intersections,through,backgrounds,decorations.pathmorphing,arrows.meta, decorations.pathreplacing}
\usepackage{subfigure}
\usepackage{latexsym}
\usepackage{appendix}
\numberwithin{equation}{section}

\newcommand{\cv}{$\mathcal{C}\mathcal{V}$ }
\newcommand{\ca}{$\mathcal{C}\mathcal{A}$ }

\newcommand{\sgn}{\text{sign}}

\begin{document}

\title{Holographic Complexity of Einstein-Maxwell-Dilaton Gravity}
\author{Brian Swingle${}^{b,c}$ and Yixu Wang${}^a$}
\address{${}^a$ Maryland Center for Fundamental Physics, \\
and Department of Physics, University of Maryland, College Park, MD 20742, USA}
\address{${}^{b}$Condensed Matter Theory Center, Maryland Center for Fundamental Physics, \\
Joint Center for Quantum Information and Computer Science, \\
and Department of Physics, University of Maryland, College Park, MD 20742, USA}
\address{${}^{c}$Kavli Institute for Theoretical Physics, Santa Barbara, CA 93106, USA}
\date{\today}

\begin{abstract}
We study the holographic complexity of Einstein-Maxwell-Dilaton gravity using the recently proposed ``complexity = volume'' and ``complexity = action" dualities. The model we consider has a ground state that is represented in the bulk via a so-called hyperscaling violating geometry. We calculate the action growth of the Wheeler-DeWitt patch of the corresponding black hole solution at non-zero temperature and find that, depending on the parameters of the theory, there is a parametric enhancement of the action growth rate relative to the conformal field theory result. We match this behavior to simple tensor network models which can capture aspects of hyperscaling violation. We also exhibit the switchback effect in complexity growth using shockwave geometries and comment on a subtlety of our action calculations when the metric is discontinuous at a null surface.
\end{abstract}

\section{Introduction}

The physics of quantum information has played a growing role in our understanding of the emergence of spacetime and gravity from non-gravitational degrees of freedom in the context of the AdS/CFT correspondence. Entanglement~\cite{Maldacena_EBH_ADS,Ryu_RT,Raamsdonk_GravityEntanglement,Swingle_AdS_MERA}, quantum error correcting codes~\cite{Harlow_AdS_QEC}, and even quantum state complexity~\cite{Susskind_Complexity} have all been used to illuminate various mysterious aspects of the emergent gravitational degrees of freedom. Here we focus on quantum state complexity and on its conjectured holographic dual \cite{Susskind_Complexity,Stanford_ComplexityShocks,Brown_CA&BH,Chapman_Complexity_Formation}.

Tensor networks have played an important role in these developments by providing a middle ground between quantum gravity and quantum information where many features of both sides can be cleanly identified and studied~\cite{Swingle_AdS_MERA,Evenbly_TNGeometry,Pastawski_PerfectTensors,Hayden_RandomTensors,Hartman_BH_interior, Molina-Vilaplana_Hartman-Maldacena,Czech_TNQuotient}. In the context of complexity, the early discussions more-or-less identified the complexity of the field theory state with the number of tensors in the minimal tensor network needed to prepare the state (up to a constant)~\cite{Susskind_Complexity,Stanford_ComplexityShocks,Brown_CA&BH}. More sophisticated definitions are now being explored in various contexts~\cite{Nielsen_ComplexityGeometry,Jefferson_FreeQFTComplexity,Chapman_FreeQFTComplexity}.

On the gravity side, complexity has been conjectured to be dual to various geometric measures, including the volume of a certain maximal slice (``complexity = volume'' or \cv) \cite{Stanford_ComplexityShocks} and the action of a certain spacetime region (``complexity = action'' or \ca) \cite{Brown_CA&BH,Brown_CA}. In particular, in the context of black hole geometries, it has been argued that the growth of the interior of the black hole is dual to the growth of the complexity of the field theory state under time evolution \cite{Hartman_BH_interior,Susskind_Complexity}. Moreover, using the action prescription it was observed that at late times a large class of black holes of equal mass have the same action growth rate and hence are conjectured to have the same complexity growth rate. This led to a conjecture that black holes complexify as rapidly as possible \cite{Brown_CA&BH,Lloyd_UltimateBound}; however, this proposal is known to be violated at least at early times~\cite{Carmi_HoloComplexityTimeDep} and in sufficiently exotic computational setups \cite{Jordan_NoBound,Brown_CA&BH}. Very recently, while this paper was being prepared, some holographic examples showing late time violations of the proposed bound were exhibited~\cite{Couch_NonCommActionGrowth,Moosa_DivergeRateHoloComplexity}. There is a growing body of work extending these results to a wider class of gravity theories and exploring other proposals~\cite{Carmi_CommentsHC, Couch_NoetherCV, Huang_HoloC&2identities, Cai_AdSBH_growthrate, Cai_ActionChargedBH,Yang_SEC&Cbound,Fu_HCnonlocal,Reynolds_CindS,Kuang_EMD, HIS_complexity&gauge, Mansoori_CG&BHT,Alishahiha_Complexity&FR, Momeni_fidelity}.

In this work, we study holographic complexity, meaning state complexity on the field theory side and its purported duals on the gravity side, in the context of a very broad class of gravitational theories known as Einstein-Maxwell-Dilaton (EMD) theories \cite{Huijse_HiddenFS}. This is a class of models that have been considered in particular in the AdS/CMT (condensed matter theory) literature as possible starting points for describing aspects of strongly interacting physics at finite charge density, e.g., as occurs in the solid state~\cite{Charmousis_EffLowTempHolo,Ogawa_HoloFS,Huijse_HiddenFS,Dong_AspectsHSV}. For our purposes, the main interesting feature of these models is that they are dual to a broader class of scale invariant (but not conformally invariant) field theories and hence have a more general tensor network representation~\cite{Evenbly_BranchingMERA_Entanglement,Swingle_AreaLawEntanglementThermo}.

Our results are as follows. Focusing on so-called hyperscaling violating solutions of the EMD theory~\cite{Huijse_HiddenFS}, we compute the rates of action growth and volume growth for finite temperature states dual to black holes as a function of the energy and the two scaling parameters describing the solutions. As we discuss in detail below, the ``dynamical exponent'' $z$ controls the relative scaling of space and time while the ``hyperscaling violation exponent'' $\theta$ relates to the effective dimensionality of space. In terms of these parameters, we find that for $z=0$, the previously obtained action growth for conformal field theories is obtained for all $\theta$. However, for $z>1$ we find that the action growth rate is enhanced by a multiplicative factor,
\beq
 \frac{\delta I}{\delta t}=2E\(1+\frac{z-1}{d-\theta}\).
\eeq
Thus these black holes violate the conjectured action growth bound even at late times.

We are able to match these results to a simple tensor network model of complexity growth in hyperscaling violating field theories. We also compare our tensor network results to the rate of volume growth and find that we need a temperature dependent length scale on the gravity side to match tensor network expectations. Finally, we study shockwave solutions and verify the existence of the switchback effect; we also point out a subtlety concerning the proper definition of action when the metric is discontinuous.

\section{Gravity model}
\subsection{Einstein-Maxwell-Dilaton theory}
The Einstein-Maxwell-Dilaton (EMD) theory is $(d+2)$ dimensional Einstein gravity sourced by a $U(1)$ gauge field (the Maxwell field) and by a scalar field $\Phi$ (the dilaton). The gauge field is coupled to the dilaton via a warping of the effective gauge coupling. The Lagrangian density can be written in terms of general functions $V$ and $Z$ via
\beq\label{lagrangian}
\CL_{EMD}=\frac{1}{2\kappa^2}\left(\CR-2(\nabla\Phi)^2-\frac{V(\Phi)}{L^2}\right)-\frac{Z(\Phi)}{4e^2}F_{\mu\nu}F^{\mu\nu}
\eeq
where ${2\kappa^2}=16\pi G$ is the gravitational constant. The action of the theory is
\beq\label{action}
I=\int d^{d+2}x\sqrt{|g|} \CL_{EMD} + ...
\eeq
where $...$ denotes additional boundary and corner terms which we specify later.

Now consider the following ansatz for the metric,
\beq\label{ansatz}
ds^2=L^2\(-f(r)dt^2+g(r)dr^2+\frac{dx_i^2}{r^2}\),
\eeq
in which the length scale $L$ reduces to the AdS radius in the conformal limit and $r$ is the emergent holographic direction. We also assume only the time component of the gauge field is non-zero,
\beq
A_t=\frac{eL}{\kappa}h(r).
\eeq

Following Ref.~\cite{Huijse_HiddenFS}, consider the situation in which the dilaton potential $V(\Phi)$ and the coupling $Z(\Phi)$ take the following asymptotic form as $\Phi\rightarrow\infty$,
\bea
Z(\Phi)=& Z_0 \exp\(\alpha \Phi\)\\
V(\Phi)=&-V_0 \exp\(\beta \Phi\),
\eea
where $\alpha,\beta$ are two positive constants. Given these forms, one solution of the EMD equations of motion is given by Eq.~\eqref{ansatz} with
\bea\label{zeroTmetric}
f(r)=&(\hat Q^{1/d}r)^{-2-2d(z-1)/(d-\theta)}V_{0}^{-1}(V_0 Z_0)^{-\frac{\theta}{d(d-\theta)}}f_0\\ \nonumber
g(r)=&\hat Q^{2/d}(\hat Q^{1/d}r)^{-2-\frac{2\theta}{d(d-\theta)}}V_{0}^{-1}(V_0 Z_0)^{-\frac{\theta}{d(d-\theta)}}g_0\\ \nonumber
h(r)=&(\hat Q^{1/d}r)^{-d-dz/(d-\theta)}h_0\\ \nonumber
e^{\Phi(r)}=&\(\hat Q^{1/d}(r/r_0)(V_0 Z_0)^{-1/2d}\)^{\frac{2d}{\alpha}(1+\frac{\theta}{d(d-\theta))})}
\eea
where the parameters are
\bea
\hat Q=&\CQ\frac{\kappa e}{L^{d-1}}\\
\theta=&\frac{d^2}{\alpha+(d-1)\beta}\\
z=&1+\frac \theta d +\frac{8(d(d-\theta)+\theta)^2}{d^2(d-\theta)\alpha^2}.
\eea

$\CQ$ is an integration constant identified as the charge density in the dual field theory, and is defined as
\beq
\CQ=-\frac{L^{d-1}}{\kappa e}\frac{h'(r) Z(\Phi(r))}{r^d\sqrt{f(r)g(r)}}.
\eeq

Given the solution in Eq.~\eqref{zeroTmetric}, we see that in order to have each term in Eq.~\eqref{ansatz} scale accordingly, the coordinates should scale as

\bea
\label{xscale} x_i\to & \lambda x_i\\
r\to& \lambda^{\frac{d-\theta}{d}}r\\
 \label{tscale} t\to& \lambda^{z}t\\
 \label{dsscale}ds\to& \lambda^{\frac{\theta}{d}}ds
\eea

From these scaling relations we know that $z$ is the ``dynamical critical exponent''. For example, in a weakly coupled theory Eqs.~\eqref{xscale} and \eqref{tscale} would require a field theory dispersion relation relating energy $\varepsilon$ to momentum $k$ going like $\varepsilon\sim k^z$. From Eq.~\eqref{dsscale} we know that the proper length in the holographic theory transforms under the above scaling transformation. This implies that the hyperscaling invariance of the boundary field theory is violated as well. So $\theta$ is the so called ``hyperscaling violation exponent''.

Further, the coefficients $g_0$ and $r_0$ can be fixed and the combination of $f_0$ and $h_0$ given by $f_0 h_0^{-2}$ can also be fixed.
\bea
g_{0}=&(z-1)^{\frac{\theta}{d-\theta}}(z+d-\theta-1)^{1+\frac{\theta}{d-\theta}}(z+d-\theta)\frac{d^2}{(d-\theta)^2}\\
r_0=&(z-1)(z+d-\theta-1)^{-1}\\
\frac{f_0}{h_0^2}=&(z-1)^{-2-\frac{2\theta}{d-\theta}}(z+d-\theta-1)^{1+\frac{\theta}{d-\theta}}(z+d-\theta)
\eea

More generally, there is an infinite class of solutions of the form Eq.~\eqref{ansatz} that correspond to a black hole geometry at non-zero temperature. The metric is modified to
\bea\label{finiteTmetric}
f_T(r) =&f(r)\(1-\(\frac{r}{r_h}\)^{d(1+z/(d-\theta))}\)\\
g_T(r)=&g(r)\(1-\(\frac{r}{r_h}\)^{d(1+z/(d-\theta))}\)^{-1}
\eea
where $r_h$ is the parameter that labels the solution which is identified with the $r$ coordinate of the horizon of the black hole. We see that the metric in Eq.~\eqref{zeroTmetric} is the zero temperature limit of this class.

The Hawking temperature of the black hole is given by the surface gravity at the horizon
\beq\label{temperature}
T=\frac{\nabla_{r}f(r)}{4\pi\sqrt{f(r)g(r)}}=\frac{d(d-\theta +z)}{4\pi(d-\theta )}\sqrt{\frac{f_0}{g_0}}r_h^{-\frac{d z}{d-\theta }} \hat Q^{-\frac{z}{d-\theta }}.
\eeq
The Bekenstein-Hawking entropy is related to the surface area of the black hole horizon and takes the form
\beq\label{bhentropy}
S=\frac{L^{d}}{4}r_{h}^{-d}\Omega^d
\eeq
where $\Omega^d$ is the regulated volumn of the $d$-dimensional hypersurface in the spatial $x_i$ directions. For later use, we can obtain the thermal energy $E$ by integrating the thermodynamic equation $dE=TdS$ with boundary condition $E|_{r_h=\infty}=0$.

\beq\label{energy}
E=\frac{d}{16\pi G}L^d\Omega^d\hat Q^{-\frac{z}{d-\theta }}\sqrt{\frac{f_0}{g_0}}r_h^{-\left(d+\frac{d z}{d-\theta }\right)}
\eeq

\subsection{Action of the Wheeler-DeWitt patch}

\begin{figure}
\centering
\begin{tikzpicture}

\coordinate  (LD) at (0,0);
\coordinate  (RD) at (8,0);
\coordinate  (LU) at (0,8);
\coordinate  (RU) at (8,8);

\coordinate [label=left:$A$] (A) at (0,4.5);
\coordinate (AD) at (4.5,0);
\coordinate [label=above:$C$](AU) at (3.5,8);
\coordinate [label=right:$B$] (B) at (8,6);
\coordinate (BD) at (2,0);
\coordinate [label=above:$D$](BU) at (6,8);

\path [name path=AAD] (A) to(AD);
\path [name path=BBD] (B) to(BD);
\path [name intersections={of=AAD and BBD, by={[label=below:$E$]E}}];

\filldraw[gray!40] (AU)--(A)--(E)--(B)--(BU);
\draw (AU)--(A)--(E)--(B)--(BU);

\draw [dashed](LD) to (RU);
\draw [dashed](LU) to (RD);
\draw (LU) to (LD);
\draw (RU) to (RD);
\draw [decorate,decoration={snake,amplitude=.4mm}] (LU) to (RU);
\draw [decorate,decoration={snake,amplitude=.4mm}] (LD) to (RD);

\end{tikzpicture}
\caption{The shaded part of the Penrose diagram shows a WDW patch of a Cauchy surface that intersects $r=\infty$ at A and B. The past and future horizon are represented by dashed lines, while the past and future singularity are represented by wave lines.}
\label{PD}
\end{figure}
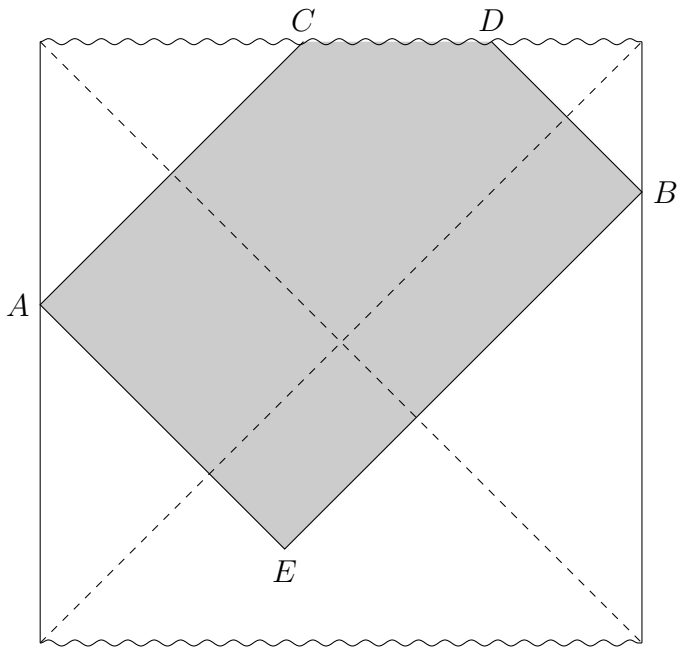

According to the \ca conjecture~\cite{Brown_CA}, the complexity of a boundary state is proportional to the classical action of a region of spacetime called the Wheeler-DeWitt patch, which is the domain of the dependence of a Cauchy surface which intersects the boundary of the spacetime at a given time. A Wheeler-DeWitt (WDW) patch in the AdS-Schwarzschild black hole spacetime is shown in Figure~\ref{PD}.

As the WDW (Figure~\ref{PD}) patch is a manifold with boundary, we shall both consider spacelike surfaces, (such as CD, a segment of the future singularity) and null surfaces (segments AC, AE, BD, BD). The extrinsic curvature of a null surface is ill-defined, so the surface action of a null surface needs detailed consideration. Furthermore, the ($d+1$)-dimensional hypersurfaces may have sharp boundaries when they intersect with each other. These joints are $d$-dimensional hypersurfaces shown as points (A, B, C \emph{etc}) in the Penrose diagram \ref{PD}.

The issue of the action for null surfaces and joints was considered in detail in Ref.~\cite{Lehner_Grav_Nullboundary}. We summarize the results that are of relevance to our calculation in Appendix~\ref{njaction}. Briefly, the total WDW patch action is given by
\beq\label{actionsum}
I=I_{bulk}+\sum_i I_{\Sigma_i}+ \sum_i I_{\CN_i} +\sum_i I_{j_i}
\eeq
in which we refer to Eq.~\eqref{action} for $I_{bulk}$, Eq.~\eqref{surfaceaction} for $I_{\Sigma_i}$, Eq.~\eqref{nullsurfaceaction} for $I_{\CN_i}$, and Eq.~\eqref{jointaction} for $I_{j_i}$.

\subsection{Action growth in the EMD theory}

\begin{figure}[tbh]
\centering
\begin{tikzpicture}

\coordinate  (LD) at (0,0);
\coordinate  (RD) at (8,0);
\coordinate  (LU) at (0,8);
\coordinate  (RU) at (8,8);
\coordinate  [label=45:$v$](vlabel) at (8.3,8.3);
\coordinate  [label=135:$u$](ulabel) at (-0.3,8.3);

\coordinate [label=255:$(u_0\text{,}v_0)A$] (A) at (0,4.5);
\coordinate (AD) at (4.5,0);
\coordinate [label=above right:$C$](AU) at (3.5,8);

\coordinate [label=right:$B(u_1\text{,}v_1)$] (B) at (8,6);
\coordinate (BD) at (2,0);
\coordinate [label=above:$D$](BU) at (6,8);

\coordinate [label=105:$A'$] (A') at (0,5.5);
\coordinate (A'D) at (5.5,0);
\coordinate [label=above left:$C'$](A'U) at (2.5,8);

\path [name path=AAD] (A) to(AD);
\path [name path=AAU] (A) to(AU);
\path [name path=BBD] (B) to(BD);
\path [name path=A'A'D] (A') to(A'D);
\path [name intersections={of=AAD and BBD, by={[label=below:$E(u_1\text{,}v_0)$]E}}];
\path [name intersections={of=A'A'D and BBD, by={[label=315:$E'$]E'}}];
\path [name intersections={of=A'A'D and AAU, by={[label=right:$F$]F}},fill];

\begin{scope}[on background layer]
\filldraw[green!15] (AU)--(A)--(E)--(B)--(BU);
\filldraw[blue!50] (A'U)--(A')--(E')--(B)--(BU);
\filldraw[blue!40!green!40] (AU)--(F)--(E')--(B)--(BU);
\end{scope}

\draw (A'U)--(A')--(E')--(B)--(BU);
\draw (AU)--(A)--(E)--(B)--(BU);

\path (A) edge  node [above]{$V_1$} node [below,sloped]{$u=u_0$}(AU);
\path (A) edge  node [above]{$V_2$} node [below,sloped]{$v=v_0$}(E);
\path (A') edge  node [above,sloped]{$u=u_0+\delta t$}(A'U);
\path (A') edge  node [above,sloped]{$v=v_0+\delta t$}(E');
\path [|<->|]($ (A.west) - (0.3,0) $) edge  node [left]{$\delta t$}($ (A'.west) - (0.3,0) $);
\path [|<->|]($ (AU.north) + (0,0.2) $) edge  node [above]{$\delta \Sigma$}($ (A'U.north) + (0,0.2) $);

\draw [dashed,-Stealth] (-0.2,-0.2) -- (vlabel);
\draw [dashed,-Stealth] (8.2,-0.2) -- (ulabel);
\draw (LU) to (LD);
\draw (RU) to (RD);
\draw [decorate,decoration={snake,amplitude=.4mm}] (LU) to (RU);
\draw [decorate,decoration={snake,amplitude=.4mm}] (LD) to (RD);

\end{tikzpicture}
\caption{ Illustration of the change of the WDW patch after evolving $t_L$ by $\delta t$. The Penrose diagram for an EMD black hole shares a similar structure to that of the AdS-Schwartzschild black hole, and the deviation does not affect the analysis hereafter.}
\label{WDWchange}
\end{figure}
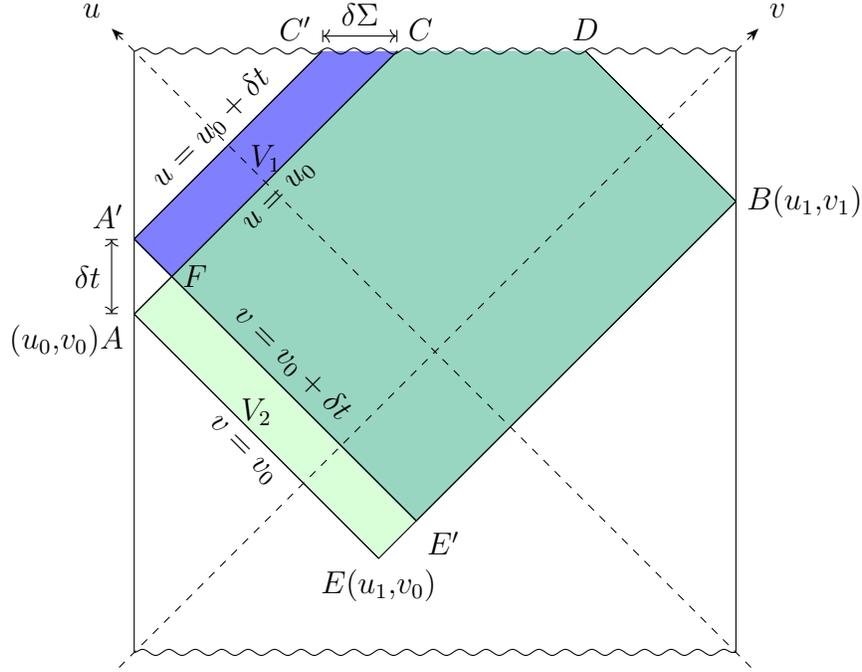

In this part we focus on the rate of change of the action as a function of time rather than on the absolute value of the action. Note that the WDW patch depends on the Cauchy surface through its intersection with the boundary at $r\to 0$, (e.g. A, B in Figure \eqref{PD}) rather than on the specific details of the Cauchy surface. Our primary interest is thus the dependence of the action of the WDW patch on the combination $(t_A+t_B)$. The combination $(t_A+t_B)$ appears because the time in the left and right spatial regions flow in opposite directions. Without loss of generality, we study the time evolution as a function of the left time $t_A = t_L$. An illustration of the change of the WDW patch when we evolve the $t_L$ by $\delta t$ is shown in Figure \ref{WDWchange}. Note that the deviation of EMD black hole metric from the AdS-Schwarzschild metric does not affect the qualitative structure of the Penrose diagram.

After analyzing the different parts that may contribute to the action growth (see Appendix~\ref{caldetail} for the details of the calculation), it turns out that only the bulk regions $V_1$ and $V_2$, the spacelike surface section $\delta\Sigma$, and the null-null surface joints $E$, $E'$ will contribute to the change of the action:
\beq\label{actionchange}
\begin{split}
\delta I&=\int_{V_1} dr dt d^{d}x_i\sqrt{|g|} \CL_{EMD}-\int_{V_2} dr dt d^{d}x_i\sqrt{|g|} \CL_{EMD}-\frac{1}{8\pi G}\int_{\delta\Sigma} dt d^{d}x_i \sqrt{|h|} K\\
&+\frac{1}{8\pi G}\int_{E'} a_{E'} \sqrt{\gamma} d^d x_i-\frac{1}{8\pi G}\int_{E} a_E \sqrt{\gamma} d^d x_i
\end{split}
\eeq

In the late time limit, $t_A\to \infty$, the null surface $BE$ lies close to the horizon and the action growth rate takes a remarkably simple form in terms of $E, \theta, z$ and $d$.
\beq\label{growthrate}
\frac{\delta I}{\delta t}=2 E \left(1+\frac{z-1}{d-\theta}\right)
\eeq
Observe that in the limit $\theta \to 0$ and $z\to 1$, \eqref{growthrate} recovers the result for AdS-Schwarzschild black holes~\cite{Brown_CA&BH}.

This indicates a significant violation of the complexity growth bound conjectured in Ref.~\cite{Brown_CA&BH} and inspired by the Lloyd's conjecture~\cite{Lloyd_UltimateBound}. However, we remind the reader that early time violations of the conjecture were already known and that some models of computation have been exhibited which also violate the conjectured bound. Nevertheless, within the confines of the \ca duality conjecture, it seems that hyperscaling violating black holes complexify much more rapidly than their conformal cousins.

\subsection{Maximal volume slice and $\CC\CV$ duality}
In the $\CC \CV$ conjecture \cite{Stanford_ComplexityShocks}, the complexity of a state $|\psi(t_L, t_R)\rangle$ is taken to be proportional to the volume of a maximal volume slice which intersects with the two $r\to 0$ boundaries at $t_L$ and $t_R$.

At late times, the maximal volume slice asymptotes to a fixed slice as $t_L\to\infty$ and $t_R\to\infty$~\cite{Stanford_ComplexityShocks}. When $t_L\to\infty$ and $t_R\to\infty$, the configuration has time translation invariance so the shape of the maximal volume slice is independent of $t_L, t_R$. It turns out that in this case the slice is a constant $r$ surface whose value $r_m$ is obtained by maximizing the measure, which amounts to finding critical points:
\beq
\partial_{r}\left(L^{d+1}\sqrt{|f(r)|}r^{-d}\right)=0.
\eeq
The solution $r=r_m$ is proportional to $r_h$,
\beq\label{h-m}
\left(\frac{r_m}{r_h}\right)^{d+dz/(d-\theta)}=2+\frac{2\theta}{d^2+dz+d\theta-2\theta}.
\eeq
Then the change rate of the volume with respect to $t_L$ or $t_R$ is given by the spatial volume of the $d$-dimensional hypersurface $V_d$ in the $x_i$ directions,
\beq
\begin{split}
\frac{\delta\CV}{\delta t}=V_d&=L^{d+1}\Omega^d r_m^{\frac{\theta }{d-\theta }-\left(\frac{d z}{d-\theta }+d\right)} \sqrt{\frac{f_0 Z_0 (d (d-\theta +z)) \hat Q^{\frac{2 \theta -2 d z}{d (d-\theta )}} \left(V_0 Z_0\right)^{\frac{1}{\theta -d}+\frac{1}{d}-1}}{d^2-d \theta +d z-2 \theta }}\\
&\sim r_h^{\frac{\theta }{d-\theta }-\left(\frac{d z}{d-\theta }+d\right)}\sim E (\hat Q^{1/d}r_h)^{\frac{\theta}{d-\theta}}\sim E T^{-\frac{\theta}{dz}}.
\end{split}
\eeq

The proportionality is obvious from the relation in Eq.~\eqref{h-m}. It is worth pointing out that in the case of non-vanishing $\theta$, $V_d$ has different dependence on $r_h$ as compared to the energy E in Eq.~\eqref{energy} and an extra dimensional scale $\hat Q$ is introduced to get the right dependence on dimensionful parameters. Thus we find that while the volume and the action both grow linearly with time at late time, the rate of growth has a qualitatively different dependence on temperature in the two cases.

\section{Tensor network model}

In this section we present two tensor network models that partially capture the complexity growth of the thermofield double state corresponding to the EMD black hole geometry. When carrying out the calculations, the following assumptions are made. Since the boundary theory has a scaling symmetry, we assume that the time evolution can be ``renormalized'' by passing it through the renormalization group (RG) circuit so that it acts only on low energy degrees of freedom~\cite{Brown_CA&BH}. This renormalization already vastly reduces the naive complexity of time evolution. Rather than using a detailed model for the thermal state at a given temperature, we instead approximate the thermal state by taking the ground state RG circuit and truncating it once the renormalized correlation length is equal to the lattice spacing. At that final thermal scale where the low energy degrees of freedom reside, we assume that the complexity growth is proportional to the energy scale of the Hamiltonian and the number of degrees of freedom.

\subsection{$(d-\theta)$-dimensional tensor networks embedded in $d$-dimensions}\label{stack}

By studying the boundary behaviour of the metric Eq.~\eqref{ansatz} with solution Eqs.~\eqref{zeroTmetric},\eqref{finiteTmetric} we know that the boundary field theory should live in $d$-dimensional space (the number of $x_i$s). Here we show that the scaling of the temperature, entropy, and complexity growth can be captured by a system composed of a direct sum of $(d-\theta)$ dimensional tensor networks. We also incorporate the dynamical critical exponent $z$ into the analysis.

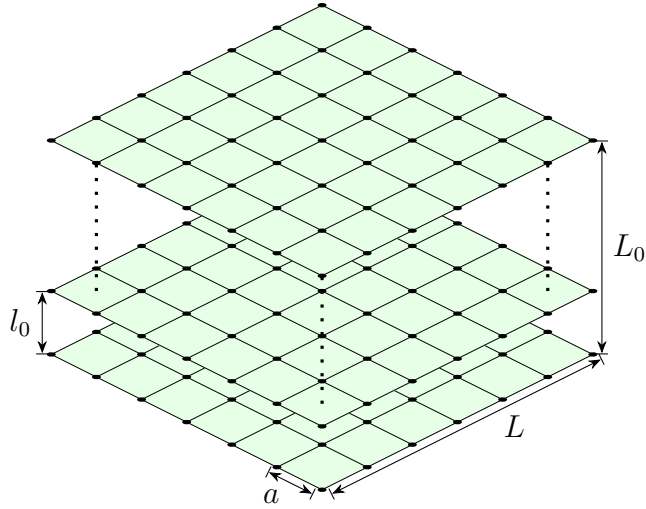
\begin{figure}[tbhp]
\centering

\begin{tikzpicture}[>=Stealth,scale=0.6]

\begin{scope}[yshift=0,yslant=0.5,xslant=-1]
\filldraw[green!10] (0,0) rectangle (6,6);
\draw[step=1, very thin] (0,0) grid (6,6);
\draw [|<->|] (-0.2,0) -- (-0.2,1) node [left,midway ,yshift=-5]{$a$};
\path [|<->|] (0,-0.2)edge  node [right,sloped, xshift=10]{$L$}(6,-0.2);

\coordinate (A) at (0,6);
\coordinate (A') at (6,0);

 \foreach \x in {0,...,6} {
    \foreach \y in {0,...,6} {
      \fill[black] (\x,\y) circle(0.07);
     }
  }
\end{scope}

\begin{scope}[yshift=40,yslant=0.5,xslant=-1]
\filldraw[green!10] (0,0) rectangle (6,6);
\draw[step=1, very thin] (0,0) grid (6,6);
\coordinate (B) at (0,6);
\coordinate (B1) at (0.5,0.5);
\coordinate (B2) at (0.5,5.5);
\coordinate (B3) at (5.5,0.5);

 \foreach \x in {0,...,6} {
    \foreach \y in {0,...,6} {
      \fill[black] (\x,\y) circle(0.07);
     }
  }
\end{scope}

\begin{scope}[yshift=135,yslant=0.5,xslant=-1]
\filldraw[green!10] (0,0) rectangle (6,6);
\draw[step=1,very thin] (0,0) grid (6,6);
\coordinate (C') at (6,0);
\coordinate (C1) at (0,0);
\coordinate (C2) at (0,5);
\coordinate (C3) at (5,0);

 \foreach \x in {0,...,6} {
    \foreach \y in {0,...,6} {
      \fill[black] (\x,\y) circle(0.07);
     }
  }
\end{scope}

\path [|<->|]($ (A.west) - (0.2,0) $) edge  node [left]{$l_0$}($ (B.west) - (0.2,0) $);
\path [|<->|]($ (A'.west) + (0.2,0) $) edge  node [right]{$L_0$}($ (C'.west) +(0.2,0) $);

\draw[loosely dotted,very thick](B1) to (C1);
\draw[loosely dotted,very thick](B2) to (C2);
\draw[loosely dotted,very thick](B3) to (C3);

\end{tikzpicture}

\caption{An illustration of a system described in Section~\ref{stack} with $d=3$, $\theta=1$. The two dimensional networks are aligned along the one dimensional vertical line. The dots are the sites of the lattice on which the Hamiltonian acts, and the ellipses denote the intermediate layers that are not drawn in the figure.}
\label{d-theta}
\end{figure}

Suppose each copy of the $(d-\theta)$ dimensional tensor network has lattice length $a$ and overall size $L$. The remaining $\theta$ dimensions are regularized to have length $L_0$, and each copy separated by a displacement of length $l_0$ along each of the $\theta$ directions. An illustration of this set up is shown in Figure~\ref{d-theta}. In total we have
\beq \label{ncopy}
N_{\text{copies}}=\(\frac{L_0}{l_0}\)^{\theta}
\eeq
As a simple example of this kind of physics, non-interacting Weyl fermions in $d=3$ spatial dimensions in the presence of a magnetic field organize at low energy into a set of $d=1$ dimensional chiral modes propagating along the magnetic field direction. There is also a similar phenomenon in holographic models~\cite{DHoker_MagBraneBH}.

We perform the RG transformation until the lattice spacing reaches some temperature dependent correlation length $\xi$ so that each site is uncorrelated with the rest. We denote the complexity generated from performing the series of RG transformation as $C_{RG}$. Now the total number of the sites in each $(d-\theta)$ dimensional network can be written as
\beq \label{nsites}
N_{\text{sites}}=\left(\frac{L}{\xi}\right)^{d-\theta}
\eeq

The dynamical critical exponent gives $\xi\sim T^{-\frac{1}{z}}$. So Eq.~\eqref{nsites} can be expressed as $N_{\text{sites}}\sim L^{d-\theta}T^{\frac{d-\theta}{z}}$. Combined with Eq.~\eqref{ncopy}, we have the total number of sites in the whole system as
\beq
N_{\text{total}}=N_{\text{copies}}\times N_{\text{sites}}\sim L^{d-\theta}T^{\frac{d-\theta}{z}} \(\frac{L_0}{l_0}\)^{\theta}
\eeq

We now compute how RG transformation acts on an infinitesimal time evolution step,
\begin{equation}
V^{\dagger}e^{- i H^{(a)} \delta t} V \approx V^{\dagger}\left(I^{(a)}-i H ^{(a)}\delta t\right)V=I^{(2a)}-i 2^{-\Delta_H}H ^{(2a)} \delta t.
\end{equation}
Here $V$ is an isometry that transforms the operators defined with lattice spacing $a$ to operators defined with lattice spacing $2a$. The superscripts denote the lattice size the operators act on. $\Delta_H$ is the scaling dimension of the Hamiltonian operator. Since the dynamical critical exponent relates scaling of time and space, we take $\Delta_H=z$. In order to renormalize to the low energy degrees of freedom at scale $\xi$, we must perform the isometry $n$ times where $n$ is
\begin{equation}\label{nexpression}
n=\log_2 \left(\frac{\xi}{a}\right).
\end{equation}

The result of applying the RG isometry $n$ times is
\begin{equation}
 V^{\dagger n} \left(I^{(a)}-i\delta t H ^{(a)}\right)V^n\approx I^{(\xi)}-i 2^{- n \Delta_H}H ^{(\xi)} \delta t = I^{(\xi)}-i \frac{T}{\Lambda_0^z}H ^{(\xi)} \delta t
\end{equation}
where $\Lambda_0\sim a^{-1}$ is a momentum scale corresponding to the inverse of the lattice length and the energy scale of $H^{(a)}$ and $H^{(\xi)}$ is $\Lambda_0^z$. Thus the combination $\frac{T}{\Lambda_0^z}H ^{(\xi)}$ behaves like a Hamiltonian with energy scale $T$ acting on $(L/\xi)^{d-\theta}$ sites. Equivalently, if we want to evolve for time $t$ at the unrenormalized scale, we need only evolve for a time $\frac{T}{\Lambda_0^z} t$ at the renormalized scale.

To sum up, the complexity of the unrenormalized state evolving for time $t$ can be identified with the complexity of the RG transformation to an uncorrelated state plus the complexity of this renormalized state evolving for time $\frac{T}{\Lambda_0} t$. Let $c$ be the complexity generated by the Hamiltonian acting on each site per infinitesimal time step. Then the total complexity of the state is

\begin{equation}\label{complexityofTN}
C\sim C_{RG}+ c L^{d-\theta}T^{\frac{d-\theta}{z}} \(\frac{L_0}{l_0}\)^{\theta} \frac{T}{\Lambda_0^z} t
\end{equation}

Refering to Eqs.~\eqref{energy} and \eqref{temperature}, we see that Eq.~\eqref{complexityofTN} indeed captures the temperature dependence of the complexity growth rate in Eq.~\eqref{growthrate} up to a multiplicative factor which depends on the details of the Hamiltonian.

\subsection{$d$-dimensional branching tensor tetwork, $s=2^{\theta}$ fixed point}

In this section we present another tensor network model which has a similar character to the first model without the explicit decomposition into non-interacting copies. We still assume that the effective dispersion relation is $\varepsilon \sim k^z$, and we require the system to be at $s=2^{\theta}$ fixed point. (citation required) The structure of the tensor network is now more elaborate, similar to a so-called branching MERA tensor network in which at each stage of the RG, spatial lengths are reduced but multiple copies of the system at the longer scale are produced. If a non-branching MERA can be understood as an isometry relating lattice space $a$ and $2a$, $|\psi^{(a)} \rangle = V |\psi^{(2a)}\rangle$, then a branching MERA with $s$ branches gives
\beq
|\psi^{(a)}\rangle = V \left[ |\psi^{(2a)}\rangle^{\otimes s} \right].
\eeq

We set up the model as before: $d$ dimensional tensor networks with lattice spacing $a$ and overall size $L$. We denote the momentum scale corresponding to the inverse lattice as $\Lambda_0$.

\begin{figure}
\centering
\begin{tikzpicture}[
layer1/.style={circle, inner sep=3mm, draw=black!70,fill=green!10},
layer2/.style={circle, inner sep=2mm, draw=black!70,fill=green!10},
layer3/.style={circle, inner sep=1.5mm, draw=black!70, fill=green!10},
brace/.style={decoration={brace, raise=10pt}, decorate,thick},
scale=1.2, >=Stealth]
\node at (0,0) [layer1,label=right:$\rho(H^{(a)})$] {};
\draw [->, thick] (0,-0.5) -- (0,-2)node [left, align=center, midway]{one RG step};

\node at (-2.25,-2.5) [layer2] (21){};
\node at (-0.75,-2.5) [layer2] (22) {};
\node at (0.75,-2.5) [layer2] (23){};
\node at (2.25,-2.5) [layer2,label=right:$\rho(H^{(2a)})^{\otimes 2^{\theta}}$] (24){};
\draw [brace] ($(21)+(0,0.1) $) -- ($(24)+(0,0.1) $);
\draw [->,  loosely dotted, thick ] (21) -- ($(21)-(0,1.6)$);
\draw [->,  loosely dotted, thick ] (22) -- ($(22)-(0,1.6)$);
\draw [->,  loosely dotted, thick ] (23) -- ($(23)-(0,1.6)$);
\draw [->,  loosely dotted, thick ] (24) -- ($(24)-(0,1.6)$);

\draw [->, thick]  (-3.5,0) -- (-3.5,-4.5) node [left, align=center, midway]{$n$ RG steps};

\node at (1.5,-4.5) [layer3] (31R) {};
\node at (2,-4.5) [layer3] {};
\node at (2.5,-4.5) [layer3] {};
\node at (3,-4.5) [layer3, label=right:$\rho(H^{(\xi)})^{\otimes 2^{n\theta}}$] (34R) {};
\node at (-1.5,-4.5) [layer3] (31L) {};
\node at (-2,-4.5) [layer3] {};
\node at (-2.5,-4.5) [layer3] {};
\node at (-3,-4.5) [layer3] (34L) {};
\draw [brace] ($(34R)-(0,0.15) $) -- ($(34L)-(0,0.15)$) node [below,midway, yshift=-10] {$2^{n\theta}$ copies} ;
\draw [loosely dotted, thick ] (-1,-4.5) -- (1,-4.5);

\end{tikzpicture}

\caption{An illustration of a branching tensor network with $\theta=2, s=4$. The circles at each layer represent copies of the thermal density matrix of the corresponding Hamiltonian. After $n=\log_2 \left(\frac{\xi}{a}\right)$ RG steps, we get $2^{n\theta}$ copies of the density matrices $\rho(H^{(\xi)})$, corresponding to the Hamiltonian on a lattice length $\xi$, where $\xi$ is the correlation length.}
\label{matrixsplitting}

\end{figure}
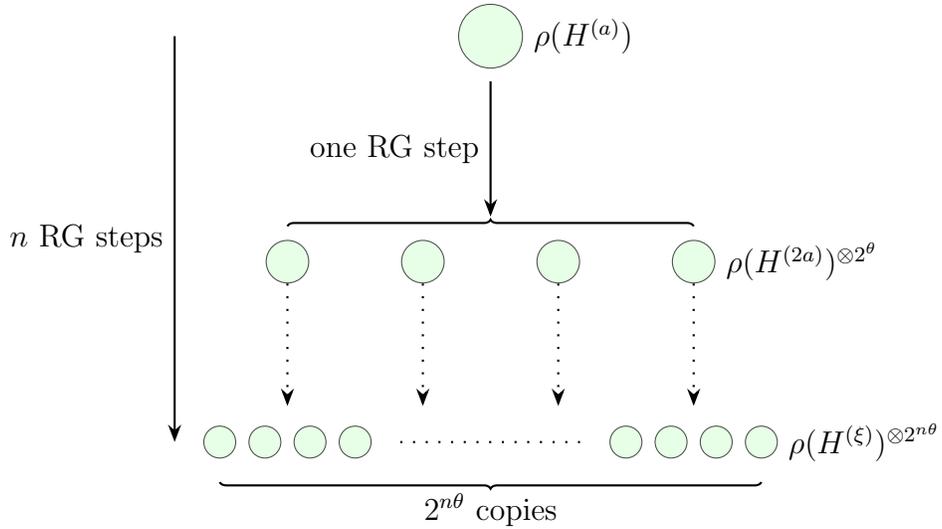

At an $s=2^{\theta}$ fixed point, the thermal density matrix $\rho(H)\propto \exp(-H/T)$ splits into a direct product of $s$ copies after one RG transformation step
\begin{equation}
\rho(H^{(a)})= V \overbrace{ \rho(H^{(2a)})\otimes \rho(H^{(2a)})\otimes...\otimes \rho(H^{(2a)})}^{2^{\theta}\text{copies}} V^\dagger
\end{equation}
An illustration of the density matrix splitting for $\theta=2$ is shown in Figure~\ref{matrixsplitting}.

We assume that after performing $n=\log_2 \left(\frac{\xi}{a}\right)$ iterations of the isometry $V$, the Hamiltonian and the state decompose into $2^{n \theta}$ disjoint copies:
\bea
 V^{\dagger n}\left(I^{(a)}-i H ^{(a)} \delta t\right)V^n=&I^{(\xi)}-i 2^{-\Delta_H n}\sum_{\ell=1}^{2^{n\theta}} H ^{(\xi)}_\ell \delta t \\
V^{\dagger n}\rho(H^{(a)}) V^{n}=&\overbrace{ \rho(H^{(\xi)})\otimes \rho(H^{(\xi)})\otimes...\otimes \rho(H^{(\xi)})}^{2^{ n\theta}\text{copies}}.
\eea
Here each $H^{(\xi)}_\ell$ is a distinct decoupled Hamiltonian acting on one of the $2^{n\theta}$ copies of $\rho(H^{(\xi)}$. The assumption of complete decoupling of the copies may be unrealistic, but we expect corrections to it will not modify the complexity growth estimate dramatically.

The counting of degrees of freedom is now similar to the previous model. For the renormalized state with lattice spacing $\xi$, the number of sites in each copy is
\beq
N_{\text{sites}}=\(\frac{L}{\xi}\)^d\sim L^d T^{\frac dz}
\eeq
The density matrix splitting effect gives
\beq
N_{\text{copies}}=2^{n\theta}=\(\frac{\xi}{a}\)^{\theta}=\frac{\Lambda_0^\theta}{T^{\frac \theta z}}
\eeq
With all the same arguments as for the previous model, our complexity estimate is
\begin{equation}\label{complexityofTNS}
C\sim C_{RG}+ c L^{d}T^{\frac{d-\theta}{z}}T \Lambda_0^{\theta -z} t
\end{equation}
which has the same temperature dependence as Eq.~\eqref{complexityofTN} and thus also captures the growth rate Eq.~\eqref{growthrate} .

\section{Shockwaves and the switchback effect}

In this section we study \ca duality for hyperscaling violating black holes in the presence of a bulk shockwave. This shockwave is dual to the insertion of a perturbation in the past of the thermofield double state. The complexity added by this perturbation can be understood in terms of the minimal quantum circuit needed to apply the Heisenberg operator $W(t_w) = e^{i H t_w} W e^{-i H t_w}$ to the thermofield double state. As discussed in Ref.~\cite{Stanford_ComplexityShocks}, we expect a partial cancellation of the forward and backward time evolutions generating $W(t_w)$, so that the total additional complexity for large $t_w$ is proportional to $2(t_w - t_*)$ with $t_*$ the scrambling time.

Here we verify that this ``switchback effect'' is also present for hyperscaling violation black holes. We do this in the context of \ca duality by explicitly evaluating the action growth of an eternal black hole perturbed by a shockwave. We find that the switchback effect is indeed reproduced and roughly matches our tensor network expectations. One subtlety which arises is how to correctly evaluate the action when the spacetime has discontinuities along null surfaces as in the shockwave geometry. We show that one way to consistently calculate the action using the ``Perimeter-style'' method of Ref.~\cite{Lehner_Grav_Nullboundary} (so that it agrees with the ``Stanford-style'' method of Ref.~\cite{Brown_CA&BH} and reproduces our physical expectations) is to open the geometry up along the null discontinuity and take a two-sided limit.

It is convenient to carry out the shockwave calculation in Kruskal-Szekeres coordinates. These coordinates can be defined throughout the eternal black hole spacetime as
\begin{equation}
\begin{split}
&U=-e^{-\frac{2\pi}{\beta}u},V=e^{\frac{2\pi}{\beta}v} \,\,\text{(right exterior region)}\\
&U=e^{-\frac{2\pi}{\beta}u},V=e^{\frac{2\pi}{\beta}v} \,\,\text{(black hole region)}\\
&U=e^{-\frac{2\pi}{\beta}u},V=-e^{\frac{2\pi}{\beta}v} \,\,\text{(left exterior region)}\\
&U=-e^{-\frac{2\pi}{\beta}u},V=-e^{\frac{2\pi}{\beta}v} \,\,\text{(white hole region)}\\
\end{split}
\end{equation}
where $\beta$ here is again the inverse temperature,
\begin{equation}
\beta=\frac{4\pi}{\partial_{\rho} f(1/\rho)|_{\rho_h}}.
\end{equation}

We set up the configuration such that the null shell is injected from the left boundary at time $t_w \rightarrow -\infty$ with infinitesimal energy $\delta\varepsilon$. When the shell arrives at the horizon, the $UU$ component of the energy-momentum tensor is exponentially blue shifted to
\beq
T_{UU}=\frac{\delta\varepsilon}{L^{d+2}} e^{2\pi |t_w|/ \beta}\delta(U).
\eeq
As stated above, the the shochwave leads to a discontinuity of the metric at the $U=0$ horizon. The discontinuity turns out to be a finite shift in the Kruskal-Szekeres variable $V$
\beq
\delta V=h\sim e^{2\pi (|t_w|-t_*)/ \beta}
\eeq
where $t_*=\frac{\beta}{2\pi} \log  \frac{L^{d}}{G_N}$. An illustration of the Penrose diagram for the shockwave geometry is shown in Figure~\ref{shockwavePenrose}. The calculation of the action of the WDW patch proceeds as before, except that we find that the discontinuity at $U=0$ must be treated specially. We show in Appendix~\ref{app:shock} the details of the action calculation and discuss how the discontinuity can be dealt with (see Section.~\eqref{discontinuity}).

The result of the calculation is
\begin{equation}
I_{\text{total}}=\(2(|t_w|-t_*)+t_L-t_R\) 2E  \left(1+\frac{z-1}{d-\theta}\right) + ...
\end{equation}
where $...$ denotes other terms that are not time-dependent. We see two principle features. First, the extra action contributed by the perturbation is proportional twice $t_w$ times the previously computed action growth rate without the shock. Second, the switchback effect is present as expected. As shown in Ref.~\cite{Roberts_ScramblingHSV}, the scrambling time in the hyperscaling violating black hole is still proportional to $\frac{\beta}{2\pi}$, i.e., they continue to maximally scramble~\cite{Maldacena_ChaosBound}. These features are also present in the tensor network models discussed above.

\section{Discussion}

In this work we studied the \ca and \cv conjectures in the context of a general class of scaling solutions to the EMD theory. We found that \ca and \cv differ in their temperature dependence for these EMD black holes. In particular, when the dynamical exponent $z$ is larger than $1$ , we found that the rate of complexity growth was enhanced relative to the $z=1$ result. We were able to match the results of the \ca calculation using simple tensor network models.

There are several interesting directions to pursue. Of course, the EMD theory considered here is not expected to be a UV complete theory of quantum gravity, so it would also be interesting to study action growth and its analogs in a more complete theory, perhaps a string theory, to at least gain some insight into the physics of the singularity. Other directions include the formulation of a more defined tensor network model and comparisons to the recent free field theory complexity calculations, perhaps in the context of the branching tensor network in Ref.~\cite{Haegeman_WaveletRigor}. Finally, given that the conjectured complexity growth rate bound in Ref.~\cite{Brown_CA&BH} is now thoroughly falsified~\cite{Couch_NonCommActionGrowth,Jordan_NoBound}, it is interesting to consider other possible bounds that might illuminate in which senses black holes are the fastest computers.

\textit{Acknowledgements:} This work is supported by the Simons Foundation as part of the It From Qubit collaboration. We thank T. Jacobson, R. Myers, S. Chapman, C. Agon, and M. Headrick for useful discussions. \textbf{Note added:} We especially thank M. Alishahiha, A. F. Astanehb, A. Mollabashi and M. R. M. Mozaffar for discussions regarding a technical error in the first version of the manuscript. They have reported independent results that agree with ours where they overlap.

\begin{appendix}
\section{Rules for calculating the action: null surfaces and joints}
We list the results that are of relevance to our calculation below. They are discussed in detail in Ref.~\cite{Lehner_Grav_Nullboundary}.

\begin{quotation}\label{njaction}

\noindent$\bullet$ For a spacelike / timelike (d+1) dimensional hypersurface, the action is given by the York-Gibbons-Hawking surface action
\beq\label{surfaceaction}
I_{YGH}=\sgn(\Sigma)\frac{1}{8\pi G}\int_{\Sigma}\sqrt{|h|} K d\Sigma
\eeq
where $\Sigma$ is the surface of interest, $h=\det h_{ab}$ is the determinent of induced metric $h_{ab}$, and $K=h^{ab}K_{ab}$ is the extrinsic curvature. If $\Sigma$ is spacelike, $\sgn(\Sigma)=1 (-1)$ if $\Sigma$ lies to the past (future) of the bulk of interest. If $\Sigma$ is timelike, then $\sgn(\Sigma)=1$.

\noindent$\bullet$ For a null surface $\CN$, the boundary action is given by
\begin{equation}\label{nullsurfaceaction}
I_{\CN}=-\sgn(\CN)\frac{1}{8\pi G}\int_{\CN}\kappa \sqrt{\gamma} d^d x_i d\lambda
\end{equation}
where $\sqrt{\gamma} d^d x_i$ is the volumn element for the d dimensional spacelike hypersurface and $\lambda$ is the parameter of the geodesic that generates the null surface $\CN$. $\sgn(\CN)=1 (-1)$ for $\CN$ in the past (future) of the bulk of interest. If we denote the vector along the null geodesic as $k^{\alpha}=\frac{\partial x^{\alpha}}{\partial \lambda}$, then the surface gravity $\kappa$ is given by
\begin{equation}\label{surfacegravity}
k^{\beta}\nabla_{\beta}k^{\alpha}=\kappa k^{\alpha}
\end{equation}
Note that if $\lambda$ is chosen to be an affine parameter of the null surface, then by definition $\kappa=0$. So when we affinely parametrize the null surface $\CN$, the corresponding boundary action $I_{\CN}=0$.

\noindent$\bullet$ For joints that come from intersection of at least one null surface, the boundary action is given by
\beq\label{jointaction}
I_{j}=\sgn(j)\frac{1}{8\pi G}\int_{j} a_j \sqrt{\gamma} d^d x_i
\eeq
If a null surface $\CN$ intersects with a spacelike surface $\CS$, and if $k$ is the null tangent vector of $\CN$ and $n$ is the normal vector of $\CS$, then the function $a_j$ is given by
\beq\label{spacenull}
a_j=\log |k\cdot n|.
\eeq

If a null surface $\CN$ intersects with another null surface $\bar\CN$, and if $k$ and $\bar k$ are the null tangent vectors of $\CN$ and $\bar\CN$ respectively, the function $a_j$ is given by
\beq\label{nullnull}
a_j=\log \left|\frac 12 k\cdot \bar k\right|.
\eeq
The rule of the sign is that $\sgn(j)=1$ if the null segment $\CN$ lies to the past (future) of the bulk of interest and the joint is at the past (future) end of the segement, and  $\sgn(j)=-1$ otherwise.
\end{quotation}

\section{Action growth in the EMD theory: calculational details} \label{caldetail}
The incoming and outgoing Eddington-Finkelstein coordiantes are convenient for the calculation. By defining
\begin{equation}
\rho=\frac 1r, v=t+\rho^{*}, u=t-\rho^{*}
\end{equation}
where $\rho^{*}$ is given by
\begin{equation}\label{rhostar}
\frac{d\rho^{*}}{d\rho}=\rho^{-2}\left(\frac{g(1/\rho)}{f(1/\rho)}\right)^{1/2},
\end{equation}
we get the variations of the metric Eq.~\eqref{ansatz}

\bea \label{urho}
ds^2=&L^2\left(-f(1/\rho)du^2-2\rho^{-2}\sqrt{f(1/\rho)g(1/\rho)}du d\rho+\rho^2 dx_i^2\right)\\\label{vrho}
ds^2=&L^2\left(-f(1/\rho)dv^2+2\rho^{-2}\sqrt{f(1/\rho)g(1/\rho)}dv d\rho+\rho^2 dx_i^2\right)\\ \label{uv}
ds^2=&L^2\left(-2f(1/\rho)du dv +\rho^2 dx_i^2\right).
\eea

The change of the action arising from evolving $t_L$ for $\delta t$ can be written
\beq\label{fullactionchange}
\delta I= \left(I_{V_1}-I_{V_2}\right)+I_{\delta\Sigma}+\left(I_{ACE}-I_{A'C'E'}\right)+\left(I_{C'}-I_{C}\right)+\left(I_{A'}-I_{A}\right)+\left(I_{E'}-I_{E}\right).
\eeq
The meaning of the terms are explained in Figure~\ref{WDWchange}.

We first show that the action given by the joints at A and C are independent of time.  The future directed unit normal vector of $CD$, at a constant $\rho$ slice, is
\beq\label{Sigmanorm}
n_{\rho}=-\frac{L (-g)^{\frac 12}}{\rho^2}
\eeq
The null surface $AC$ is determined by a scalar function $\Phi(u,\rho,x_i)=u-\text{const.}=0$. The future directed normal null vector of $AC$ is given by $k_{\alpha}=-\partial_{\alpha}\Phi$. The non-vanishing component is $k_u=1$.  We appeal to the metric Eq.~\eqref{urho} and use the formula Eq.~\eqref{jointaction} with Eq.~\eqref{spacenull}:
\beq
I_C=-\frac{1}{8\pi G}\int_{C} \log |k\cdot n| \sqrt{\gamma} d^d x_i=\frac{1}{8\pi G} L^d \Omega^d \(\rho^d \log\(L(-f)^{\frac 12}\)\)\Big{|}_{\rho\to 0}.
\eeq
This term may be not well defined when $\rho\to 0$, but it is independent of $u$, hence invariant under time translation. So $\left(I_{C'}-I_{C}\right)=0$.

By defining $\bar k$ such that the non-vanishing component is $\bar k_{v}=-1$, with metric Eq.~\eqref{uv} and formulas \eqref{jointaction} and \eqref{nullnull}, we can show that
\beq
I_A=\frac{1}{8\pi G}\int_{A} \log |\frac 12 k\cdot \bar k| \sqrt{\gamma} d^d x_i=-\frac{1}{8\pi G} L^d \Omega^d \(\rho^d \log\(- 2 L^2 f\)\)\Big{|}_{\rho\to \infty}.
\eeq
By the same argument as before, $I_A$ depends only on the combination of $u-v$, hence it is invariant under time translation. So $\left(I_{A'}-I_{A}\right)=0$.

For the null surface action $I_{CAE}$ with the null normal vectors chosen to be $k_u=1$ for $CA$ and $\bar k_v=1$ for $AE$, we can calculate explicitly that
\beq
k^{\beta}\nabla_{\beta}k^{\alpha}=0, \bar k^{\beta}\nabla_{\beta}\bar k^{\alpha}=0.
\eeq
Thus $I_{CAE}=0$ since $\kappa=0$ on both $AE$ and $CA$. So also is $I_{C'A'E'}=0$.

To summarize so far, the action change in Eq.~\eqref{fullactionchange} has been reduced to Eq.~\eqref{actionchange} as claimed. We now turn to the different parts in Eq.~\eqref{actionchange}.

\noindent$\bullet$ The bulk action $\left(I_{V_1}-I_{V_2}\right)$
\begin{equation}
I_{V_1}=\frac{1}{16\pi G}\int^{u_0+\delta t}_{u_0}du\int^{\rho(u,v_0+\delta t)}_{0}d\rho\int d^{d}x_i L^{d+2}\rho^{d-2}\sqrt{f(1/\rho)g(1/\rho)} \CL_{EMD}
\end{equation}
and
\begin{equation}
-I_{V_2}=-\frac{1}{16\pi G}\int^{v_0+\delta t}_{v_0}dv\int^{\rho(u_0,v)}_{\rho(u_1,v)}d\rho\int d^{d}x_i L^{d+2}\rho^{d-2}\sqrt{f(1/\rho)g(1/\rho)}\CL_{EMD}.
\end{equation}

By defining
\begin{equation}\label{Ffunction}
F(\rho)=\int d\rho L^{d+2}\rho^{d-2}\sqrt{f(1/\rho)g(1/\rho)}(2\kappa^2) \CL_{EMD}
\end{equation}
and changing variable $u=u_0+v_0+\delta t-v$, it turns out that
\begin{equation}\label{changebulk}
I_{V_1}-I_{V_2}=\Omega^d \int^{v_0+\delta t}_{v_0}dv F\left(\rho(u_1,v)\right)\approx\Omega^d \delta t \left(F\left(\rho(u_1,v_0)\right)-F(0)\right).
\end{equation}

The result is
\begin{equation}\label{bulkchangevalue}
\begin{split}
I_{V_1}-I_{V_2}&=-\frac{1}{8\pi G}\Omega^d \delta tL^d \hat Q^{-\frac{z}{d-\theta }}\rho(u_1,v_0)^{d+dz/(d-\theta)}\sqrt{\frac{f_0}{g_0}}
\end{split}
\end{equation}

\noindent$\bullet$ The spacelike surface action $I_{\delta\Sigma}$ is
\beq
I_{\delta\Sigma}=-\frac{1}{8\pi G}\int_{\Sigma}\sqrt{|h|} K dt d^dx_i
\eeq
where the extrinsic curvature $K$ is given by
\beq
K=\nabla_{\alpha}n^{\alpha}
\eeq
and the future directed normal vector $n^{\alpha}$ is defined in \eqref{Sigmanorm}.

The result is
\begin{equation}\label{boundarychangevalue}
\begin{split}
I_{\delta\Sigma}&=\frac{1}{16\pi G}\Omega^d \delta t L^d \rho^{d+2}\sqrt{\frac{-f(1/\rho)}{-g(1/\rho)}}\partial_{\rho}\left(\log (L^{d+1}(-f)^{1/2}\rho^d) \right)\Big{|}_{\rho=0}\\
&=\frac{1}{16\pi G}\Omega^d \delta t L^d \frac{d^2+dz-d\theta-2\theta}{d-\theta}\rho_h^{d+dz/(d-\theta)}\sqrt{\frac{f_0}{g_0}}\hat Q^{-z/(d-\theta)}
\end{split}
\end{equation}

\noindent$\bullet$ The joint action $\left(I_{E'}-I_{E}\right)$ is

\begin{equation}
\begin{split}
&\frac{1}{8\pi G}\int_{E'}a dS-\frac{1}{8\pi G}\int_{E}a dS\approx -\Omega^d \delta tL^d \rho^2\left(\frac{f(1/\rho)}{g(1/\rho)}\right)^{1/2}\partial_{\rho}\left( \rho^d\log\left(-L^2 f(1/\rho)\right) \right)\Big{|}_{\rho(u_1,v_0)}\\
&=\frac{1}{16\pi G}\Omega^d \delta t L^d \frac{d^2-dz-d\theta+2\theta}{d-\theta}\rho_h^{d+dz/(d-\theta)}\sqrt{\frac{f_0}{g_0}}\hat Q^{-z/(d-\theta)}\\
&-\frac{1}{16\pi G}\frac{\Omega^d \delta t L^d }{d-\theta}\rho^{d+dz/(d-\theta)}\sqrt{\frac{f_0}{g_0}}\hat Q^{-z/(d-\theta)}\\
&\times\left(-2dz+2\theta +d(d-\theta)\left(-1+(\frac{\rho_h}{\rho})^{d+dz/(d-\theta)}\right)\log(-L^2 f(1/\rho) \right)\Big{|}_{\rho(u_1,v_0)}
\end{split}
\end{equation}
We note when $\rho(u_1,v_0)\rightarrow \rho_h$, this term approaches
\begin{equation}\label{jointchangevalue}
\lim_{\rho\rightarrow \rho_h}\left(\frac{1}{8\pi G}\int_{E'}a dS-\frac{1}{8\pi G}\int_{E}a dS\right)=\frac{1}{16\pi G}\Omega^d \delta t L^d \frac{d^2+dz-d\theta}{d-\theta}\rho_h^{d+dz/(d-\theta)}\sqrt{\frac{f_0}{g_0}}\hat Q^{-z/(d-\theta)}.
\end{equation}

Combining Eqs.~\eqref{bulkchangevalue}, \eqref{boundarychangevalue}, \eqref{jointchangevalue}, and using Eq.~\eqref{energy}, we recover the result in Eq.~\eqref{growthrate}.

\section{Action of the shockwave geometry in EMD theory: calculational details}
\label{app:shock}

In this appendix we present the calculation of the action in a geometry perturbed by a spherically symmetric null shell falling into the blackhole. The null shell sets of a shockwave whose physical manifestation is a null shift along the shockwave. For the convenience of the calculation hereafter, we introduce the metric after changing the $u$ coordinate in Eq.~\eqref{urho} to the Kruskal-Szekeres variable $U$,
\begin{equation}
ds^2=L^2\left(-f(1/\rho)\frac{\beta^2 dU^2}{4\pi^2U^2}+2\rho^{-2}\sqrt{f(1/\rho)g(1/\rho)}\frac{\beta dU}{2\pi U} d\rho+\rho^2 dx_i^2\right).
\end{equation}
Recall that $\beta$ is chosen such that the metric is non-singular at the horizon,
\begin{equation}
\beta=\frac{4\pi}{\partial_{\rho} f(1/\rho)}\Big{|}_{\rho_h}.
\end{equation}

We set up the configuration such that the null shell is injected from the left boundary at time $t_w \rightarrow -\infty$ with infinitesimal energy $\delta\varepsilon$. When the shell arrives at the horizon, the $UU$ component of the energy-momentum tensor is exponentially blue shifted to
\beq
T_{UU}=\frac{\delta\varepsilon}{L^{d+2}} e^{2\pi |t_w|/ \beta}\delta(U).
\eeq
The shochwave leads to a discontinuity of the metric at the $U=0$ horizon if it is injected from the left boundary. The discontinuity turns out to be a finite shift in the Kruskal-Szekeres variable $V$
\beq
\delta V=h\sim e^{2\pi (|t_w|-t_*)/ \beta}
\eeq
where $t_*=\frac{\beta}{2\pi} \log  \frac{L^{d}}{G_N}$. An illustration of the Penrose diagram for the shockwave geometry is shown in Figure~ \ref{shockwavePenrose} . We will show in Section~\eqref{discontinuity} that the discontinuity across $U=0$ crucially contributes to the time dependence of action.

A very distinct feature of the shockwave metric is that the WDW patch can intersect with past and future singularities simultaneously, while in the unperturbed metric the patch can only touch both or intersect either past or future singularity. The condition that the patch intersects with the past singularity is
\begin{equation}
(U_0^{-1}+h)V_0^{-1}=e^{\frac{4\pi}{\beta}\rho^{*}(\rho_J)}\geq 1
\end{equation}
where $\rho_J$ is the radius of the point J where null boundaries of the past domain of dependence intersect. We can relate the $U_0, V_0$ to the time on the left and right boundary via

\begin{equation}\label{boundaryrelation}
U_0=e^{\frac{2\pi}{\beta}t_L},V_0=e^{\frac{2\pi}{\beta}t_R}.
\end{equation}

We would like to calculate the time-dependent (boundary condition dependent) part of the WDW patch action. The time-dependent bulk parts are colored in blue while the independent parts are colored in green in Figure~\ref{shockwavePenrose}. Besides the action contributed by the discontinuity Appendix~\ref{discontinuity}), the time dependence of the boundary and joint action are the same as discussed in Appendix~\ref{caldetail}. To be specific, in Figure~\ref{nonintersect}, only the action of the boundary $CD$ and the joint $J$ are considered. In Figure~\ref{intersect}, only the action of the boundaries $CD$ and $MN$ are considered.

\subsection{Contribution from the shockwave discontinuity }\label{discontinuity}

\begin{figure}[tbh]
\centering
\begin{tikzpicture}[scale=0.8]
\tikzstyle{every node}=[font=\fontsize{10}{0}]

\coordinate  (LD) at (0,0);
\coordinate  (RU) at (8,8);

\coordinate  [label=above:$E'$](EU) at (1.8,2.2);
\coordinate  [label=right:$E''$](ED) at (2.2,1.8);
\coordinate  [label=left:$G'$](GU) at (5.8,6.2);
\coordinate  [label=below:$G''$](GD) at (6.2,5.8);
\coordinate  [label=below:$O$](O) at (2.5,2.5);
\coordinate  [label=below:$O'$](O') at (4.5,4.5);

\path (EU) edge node [above,sloped]{$U=\epsilon$}(GU);
\path (ED) edge node [below,sloped]{$U=-\epsilon$}(GD);
\path(EU) edge node [below,sloped]{$V=-U_0^{-1}$} ($(EU)+(-2,2)$);
\path(GD) edge node [above,sloped]{$V=V_0$} ($(GD)+(2,-2)$);
\path(ED) edge node [below,sloped]{$V=-U_0^{-1}-h$} ($(ED)+(2,-2)$);
\path(GU) edge node [above,sloped]{$V=V_0+h$} ($(GU)+(-2,2)$);
\draw [dashed,-Stealth] (LD) -- (RU);
\draw (EU) to (GU);
\draw (ED) to (GD);
\draw (EU) to ($(EU)+(-2,2)$);
\draw (GU) to ($(GU)+(-2,2)$);
\draw (ED) to ($(ED)+(2,-2)$);
\draw (GD) to ($(GD)+(2,-2)$);
\draw [dashed](O) to ($(O)+(-3.5,3.5)$);
\draw [dashed](O') to ($(O')-(-3.5,3.5)$);
\end{tikzpicture}
\caption{Illustration of the treatment of the discontinuity along null segement $EG$. We take the limit $\epsilon \to 0$ so the null segement $E'G'$ and $E''G''$ approach to $EG$ in the same way.}
\label{illustrationdiscontinuity}
\end{figure}
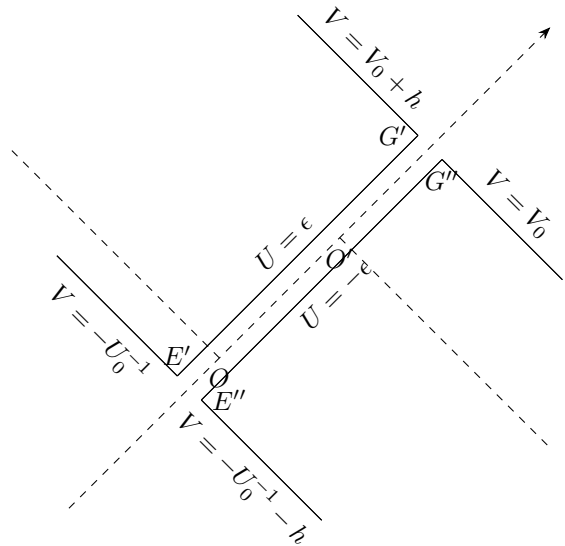

We divide the total bulk into four subregions which are colored in Figure~\ref{shockwavePenrose}, thus introducing three segements as internal boundaries: $FO, O'H, EG$. Since the metric around $FO$ and $O'H$ is continuous, the induced null surfaces and joints will have identical action up to a sign on each side and cancel with each other. However, the metric is discontinuous along $EG$, so we should not expect the action on both sides to add up to zero. We calculate the effect of the discontinuity by comparing the two null surfaces $U=\epsilon$ and $U=-\epsilon$ which approach to $EG$ in the same way when $\epsilon\to 0$ (as in Cauchy principal value integration). Figure~\ref{illustrationdiscontinuity} illustrates the treatment.

Using the Kruskal coordinate $U,V$ and Eq.~\eqref{nullsurfaceaction}, it is a straightforward calculation to obtain
\beq
I_{E'G'}=I_{E''G''}=0.
\eeq
Using Eq.~\eqref{jointaction}, we have
\beq\label{jointdifference}
\begin{split}
&I_{E'}+I_{E''}+I_{G''}+I_{G'}
\\=&\frac{\Omega^d L^d}{8\pi G}\(H^{+}(\rho_{E'})+H^{+}(\rho_{G''})-H^{-}(\rho_{E''})-H^{-}(\rho_{G'})\)
\end{split}
\eeq
where $H^{\pm}(\rho)\equiv \rho^d\log(\pm L^2f(1/\rho))$ as a short-hand notation. The $(+)$ sign applies for joints outside the future or past horizon while $(-)$ sign applies for those inside the horizon.

As $\epsilon\to 0$, the radius $\rho \to \rho_h$ for all of the four joints. Expanding $(\rho-\rho_h)$ for $H^{\pm}(\rho)$ we get
\beq\label{hexpansion}
H^{\pm}(\rho)=\rho^d_h\log(\pm(\rho-\rho_h))+c_1+O(\rho-\rho_h),
\eeq
where $c_1$ is a constant depending only on $d,z,\theta$ and is same for both $(+)$ and $(-)$ case.

On the other hand, recalling Eq.~\eqref{rhostar}, we can perform the integral and expand in $(\rho-\rho_h)$:

\beq\label{expandrhostar}
\rho^{*}(\rho)= \frac{d-\theta}{d(d-\theta+z)} Q^{\frac{z}{d-\theta }}\sqrt{\frac{g_0}{f_0}}\rho_h^{\frac{d z}{d-\theta }} \log(\pm(\rho-\rho_h))+c_2 +O(\rho-\rho_h)
\eeq
where $\pm$ sign has the same structure as $H^{\pm}(\rho)$ and $c_2$ is another constant that depends only on $d,z,\theta$ and is same for both $(+)$ and $(-)$ case.

Using the definition of the Kruskal coordinates,
\beq
UV=e^{-\frac{4\pi}{\beta}\rho^*}(\text{inside the horizon}), UV=-e^{-\frac{4\pi}{\beta}\rho^*}(\text{outside the horizon}),
\eeq
we can write
\beq\label{rhostarvalue}
\rho^{*}(\rho_E)=-\frac{\beta}{4\pi}\log(\epsilon U_0^{-1})
\eeq
and similar expressions for $E'',G',G''$.

Combining Eq.~\eqref{rhostarvalue} with Eqs.~\eqref{expandrhostar}, \eqref{hexpansion}, and using Eq.~\eqref{energy} for the energy, we can get an expression for Eq.~\eqref{jointdifference} in terms of a power series of $(\rho-\rho_h)$
\beq
\begin{split}
&I_{E'}+I_{E''}+I_{G''}+I_{G'}\\
=&\frac{\beta}{2\pi}\left(\log(1+hV_0^{-1})+\log(1+hU_0)\right) E \(1+\frac{z}{d-\theta}\)\\
+&\sum_{i=E',E'',G',G''} O(\rho_{i}-\rho_h)
\end{split}
\eeq
in which the $\log \epsilon$ terms precisely cancel with each other provided the four joints approach to the $U=0$ horizon in the same way and the higher order terms in the last line will vanish in the limit $\epsilon\to 0$.

So we conclude that the discontinuity of the metric will contribute to the total action by
\beq\label{discontinuityaction}
I_{\text{discontinuity}}\equiv I_{E'}+I_{E''}+I_{G''}+I_{G'}=\frac{\beta}{2\pi}\left(\log(1+hV_0^{-1})+\log(1+hU_0)\right) E \(1+\frac{z}{d-\theta}\).
\eeq

\subsection{Contribution of Figure~\ref{nonintersect}}
The two bulk regions behind the future and past horizon also contribute to the time dependence of the action:

\begin{equation}\label{futurehorizonbulk}
\begin{split}
I_{CFOGD}&=\int_0^{\rho_h}d\rho\int^{U_0}_{\frac{e^{\frac{4\pi}{\beta}\rho^{*}(\rho)}}{V_0+h}}\frac{\beta}{2\pi}\frac{dU}{U}\int d^{d}x_i L^{d+2}\rho^{d-2}\sqrt{f(1/\rho)g(1/\rho)}(2\kappa^2) \CL_{EMD}\\
&=\frac{\beta}{2\pi}\log(U_0(V_0+h))\Omega^d(F(\rho_h)-F(0))\\&-2\Omega^d \int_0^{\rho_h}d\rho L^{d+2}\rho^{d-2}\sqrt{f(1/\rho)g(1/\rho)}(2\kappa^2)\rho^{*}(\rho) \CL_{EMD}
\end{split}
\end{equation}

\begin{equation}
\begin{split}
I_{EJHO'}&=\int_{\rho_0}^{\rho_h}d\rho\int_{-V_0^{-1}}^{-\frac{e^{\frac{4\pi}{\beta}\rho^{*}(\rho)}}{U_0^{-1}+h}}\frac{\beta}{2\pi}-\frac{dU}{U}\int d^{d}x_i L^{d+2}\rho^{d-2}\sqrt{f(1/\rho)g(1/\rho)}(2\kappa^2) \CL_{EMD}\\
&=\frac{\beta}{2\pi}\log((U_0^{-1}+h)V_0^{-1}))\Omega^d(F(\rho_h)-F(\rho_J))\\&-2\Omega^d \int_{\rho_J}^{\rho_h}d\rho L^{d+2}\rho^{d-2}\sqrt{f(1/\rho)g(1/\rho)}(2\kappa^2)\rho^{*}(\rho) \CL_{EMD}
\end{split}
\end{equation}

where the function $F(\rho)$ is defined in Eq.~\eqref{Ffunction}. Note that the last line of both calculations are of no interest here since they are time independent.

We refer to Eq.~\eqref{surfaceaction} for the calculation of the boundary action
\begin{equation}\label{futurehorizonboundary}
\begin{split}
I_{CD}&=-2\int d\Sigma K=2\Omega^d \int^{U_0}_{1/(V_0+h)}\frac{\beta dU}{2\pi U} L^d \rho^{d+2}\sqrt{\frac{-f(1/\rho)}{-g(1/\rho)}}\partial_{\rho}\left(\log (L^{d+1}(-f)^{1/2}\rho^d) \right)\Big{|}_{\rho=0}\\
&=\frac{\beta}{2\pi}\log(U_0(V_0+h))\Omega^d L^d \frac{d^2+dz-d\theta-2\theta}{d-\theta}\rho_h^{d+dz/(d-\theta)}\sqrt{\frac{f_0}{g_0}}\hat Q^{-z/(d-\theta)}\\
\end{split}
\end{equation}

The joint action is obtained from Eq.~\eqref{jointaction}
\begin{equation}
I_{J}=2\int_{B^{\prime}}a dS=-2\Omega^d L^d \rho^d\log(-L^2 f(1/\rho))\Big{|}_{\rho_J}
\end{equation}

It is cumbersome to sum up all the parts directly, but it is easy to show that in the limit $h \to 0$ and $\rho_J \to \rho_h$, summing up all the parts gives
\beq
I_{\text{total}}=2E(t_L+t_R) \left(1+\frac{z-1}{d-\theta}\right),
\eeq
which recovers Eq.~\eqref{growthrate} when performing either a $t_L$ or $t_R$ derivative.

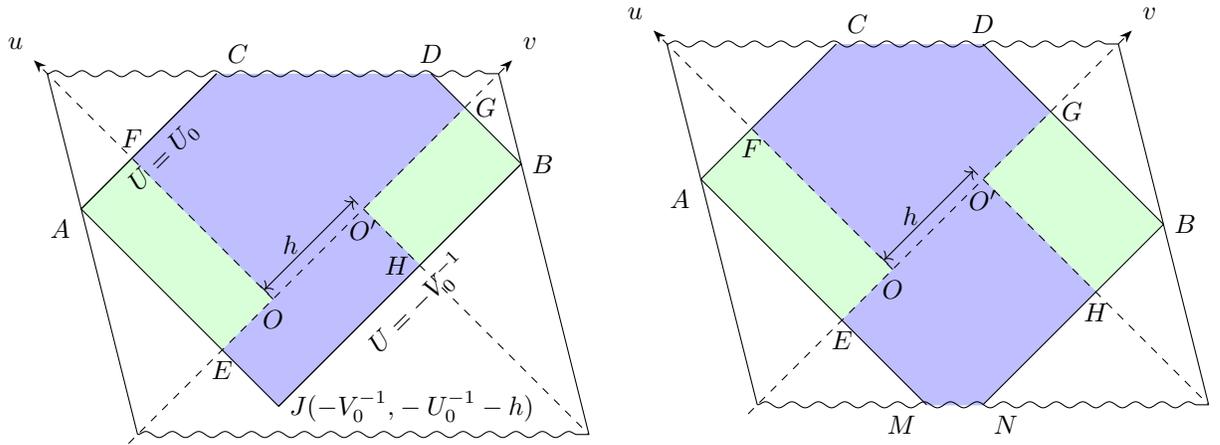
\begin{figure}[htb]

\subfigure{
\begin{tikzpicture}[scale=0.6]
\tikzstyle{every node}=[font=\fontsize{10}{0}]

\coordinate  (LD) at (0,0);
\coordinate  (RD) at (10,0);
\coordinate  (CRD) at (2,8);
\coordinate  (LU) at (-2,8);
\coordinate  (CLU) at (6,0);
\coordinate  (RU) at (8,8);

\coordinate  [label=45:$v$](vlabel) at (8.3,8.3);
\coordinate  [label=135:$u$](ulabel) at (-2.3,8.3);

\coordinate [label=255:$A$] (A) at (-1.25,5);
\coordinate (AD) at (3.75,0);
\coordinate [label=above right:$C$](AU) at (1.75,8);

\coordinate [label=right:$B$] (B) at (8.5,6);
\coordinate (BD) at (2.5,0);
\coordinate [label=above:$D$](BU) at (6.5,8);

\path [name path=AAD] (A) to(AD);
\path [name path=AAU] (A) to(AU);
\path [name path=BBD] (B) to(BD);
\path [name path=BBU] (B) to(BU);
\path [name path=uaxisU] (LU) to(CLU);
\path [name path=uaxisD] (RD) to(CRD);
\path [dashed,-Stealth,draw, name path=vaxis] (-0.2,-0.2) -- (vlabel);
\path [name intersections={of=AAD and BBD, by={[label=right:$J(-V_0^{-1}\text{,}-U_0^{-1}-h)$]J}}];
\path [name intersections={of=vaxis and uaxisU, by={[label=below:$O$]O}}];
\path [name intersections={of=AAU and uaxisU, by={[label=above:$F$]F}}];
\path [name intersections={of=vaxis and uaxisD, by={[label=below:$O'$]O'}}];
\path [name intersections={of=BBD and uaxisD, by={[label=left:$H$]H}}];
\path [name intersections={of=vaxis and AAD, by={[label=below:$E$]E}}];
\path [name intersections={of=vaxis and BBU, by={[label=right:$G$]G}}];

\path (A) edge node [below,sloped]{$U=U_0$}(AU);

\path (B) edge node [below,sloped]{$U=-V_0^{-1}$}(J);

\begin{scope}[on background layer]
\filldraw[green!15] (AU)--(A)--(J)--(B)--(BU) ;
\filldraw[blue!25] (AU)--(F)--(O)--(G)--(BU);
\filldraw[blue!25] (E)--(J)--(H)--(O');
\draw (AU)--(A)--(J)--(B)--(BU);
\end{scope}

\path [|<->|]($ (O) + (-0.2,0.2) $) edge  node [left,midway]{$h$}($ (O') + (-0.2,0.2) $);
\draw [dashed,-Stealth] (O) -- (ulabel);
\draw [dashed] (O') -- (RD);
\draw (LU) to (LD);
\draw (RU) to (RD);
\draw [decorate,decoration={snake,amplitude=.4mm}] (LU) to (RU);
\draw [decorate,decoration={snake,amplitude=.4mm}] (LD) to (RD);
\end{tikzpicture}
\label{nonintersect}
}
\subfigure{
\begin{tikzpicture}[scale=0.6]
\tikzstyle{every node}=[font=\fontsize{10}{0}]
\coordinate  (LD) at (0,0);
\coordinate  (RD) at (10,0);
\coordinate  (CRD) at (2,8);
\coordinate  (LU) at (-2,8);
\coordinate  (CLU) at (6,0);
\coordinate  (RU) at (8,8);
\coordinate  [label=45:$v$](vlabel) at (8.3,8.3);
\coordinate  [label=135:$u$](ulabel) at (-2.3,8.3);

\coordinate [label=255:$A$] (A) at (-1.25,5);
\coordinate [label=225:$M$](AD) at (3.75,0);
\coordinate [label=above right:$C$](AU) at (1.75,8);

\coordinate [label=right:$B$] (B) at (9,4);
\coordinate [label=315:$N$] (BD) at (5,0);
\coordinate [label=above:$D$](BU) at (5,8);

\path [name path=AAD] (A) to(AD);
\path [name path=AAU] (A) to(AU);
\path [name path=BBD] (B) to(BD);
\path [name path=BBU] (B) to(BU);
\path [name path=uaxisU] (LU) to(CLU);
\path [name path=uaxisD] (RD) to(CRD);
\path [dashed,-Stealth,draw, name path=vaxis] (-0.2,-0.2) -- (vlabel);

\path [name intersections={of=vaxis and uaxisU, by={[label=below:$O$]O}}];
\path [name intersections={of=AAU and uaxisU, by={[label=below:$F$]F}}];
\path [name intersections={of=vaxis and uaxisD, by={[label=below:$O'$]O'}}];
\path [name intersections={of=BBD and uaxisD, by={[label=below:$H$]H}}];
\path [name intersections={of=vaxis and AAD, by={[label=below:$E$]E}}];
\path [name intersections={of=vaxis and BBU, by={[label=right:$G$]G}}];
\begin{scope}[on background layer]
\filldraw[green!15] (AU)--(A)--(AD)--(BD)--(B)--(BU);
\filldraw[blue!25] (AU)--(F)--(O)--(G)--(BU);
\filldraw[blue!25] (E)--(AD)--(BD)--(H)--(O');
\draw (AU)--(A)--(AD) (BD)--(B)--(BU);
\end{scope}

\path [|<->|]($ (O) + (-0.2,0.2) $) edge  node [left,midway]{$h$}($ (O') + (-0.2,0.2) $);
\draw [dashed,-Stealth] (O) -- (ulabel);
\draw [dashed] (O') -- (RD);
\draw (LU) to (LD);
\draw (RU) to (RD);
\draw [decorate,decoration={snake,amplitude=.4mm}] (LU) to (RU);
\draw [decorate,decoration={snake,amplitude=.4mm}] (LD) to (RD);
\end{tikzpicture}
\label{intersect}
}

\caption{Illustration of the Penrose diagram for the shockwave geometry. Left (\ref{nonintersect}): the patch only intersects with future singularity; Right (\ref{intersect}): the patch intersects with both singularities. The bulk regions whose actions are time-independent are colored in green while those whose actions are time-dependent are colored in blue.}
\label{shockwavePenrose}
\end{figure}

\subsection{Contribution of Figure~\ref{intersect}}
In this case, the bulk regions behind the future and past horizon still contribute to the time-dependent action. The one behind the future horizon has action identical to Eq.~\eqref{futurehorizonbulk}, while the one behind the past horizon has action

\begin{equation}\label{pasthorizonbulk}
\begin{split}
I_{EMNHO'}&=\int_{0}^{\rho_h}d\rho\int_{-V_0^{-1}}^{-\frac{e^{\frac{4\pi}{\beta}\rho^{*}(\rho)}}{U_0^{-1}+h}}\frac{\beta}{2\pi}\left(-\frac{dU}{U}\right)\int d^{d}x_i L^{d+2}\rho^{d-2}\sqrt{f(1/\rho)g(1/\rho)}(2\kappa^2) \CL_{EMD}\\
&=\frac{\beta}{2\pi}\log((U_0^{-1}+h)V_0^{-1}))\Omega^d(F(\rho_h)-F(0))\\&-2\Omega^d \int_{0}^{\rho_h}d\rho L^{d+2}\rho^{d-2}\sqrt{f(1/\rho)g(1/\rho)}(2\kappa^2)\rho^{*}(\rho) L_{EMD}.
\end{split}
\end{equation}

There is no longer a joint action that is time dependent. However, we should take the boundary action of the spacelike segment $MN$ into account:

\begin{equation}\label{pasthorizonboundary}
\begin{split}
I_{MN}&=2\Omega^d \int_{-V_0^{-1}}^{-\frac{1}{U_0^{-1}+h}}\frac{\beta}{2\pi}\left(-\frac{dU}{U}\right) L^d \rho^{d+2}\sqrt{\frac{-f(1/\rho)}{-g(1/\rho)}}\partial_{\rho}\left(\log (L^{d+1}(-f)^{1/2}\rho^d) \right)\Big{|}_{\rho=0}\\
&=\frac{\beta}{2\pi}\log((U_0^{-1}+h)V_0^{-1}))\Omega^d L^d \frac{d^2+dz-d\theta-2\theta}{d-\theta}\rho_h^{d+dz/(d-\theta)}\sqrt{\frac{f_0}{g_0}}\hat Q^{-z/(d-\theta)}.\\
\end{split}
\end{equation}

The total action is obtained by summing up Eqs.~\eqref{futurehorizonbulk}, \eqref{futurehorizonboundary}, \eqref{pasthorizonbulk}, \eqref{pasthorizonboundary}, and \eqref{discontinuityaction}:

\begin{equation}
I_{\text{total}}=\frac{\beta}{2\pi}\left(\log(1+hV_0^{-1})+\log(1+hU_0)\right) 2E \left(1+\frac{z-1}{d-\theta}\right).
\end{equation}

Refering to Eq.~\eqref{boundaryrelation}, if $hV_0^{-1}\ll 1$ and $hU_0 \ll 1$, which means $t_R \gg |t_w|-t_*$ and $|t_w|-t_*\ll -t_L $, the action almost vanishes because the parts behind the past and future horizon almost cancel with each other. However, if $hV_0^{-1}\gg 1$ and $hU_0 \gg 1$, the action reduces to

\begin{equation}\label{switchback}
I_{\text{total}}=\(2(|t_w|-t_*)+t_L-t_R\) 2E \left(1+\frac{z-1}{d-\theta}\right).
\end{equation}

Eq.~\eqref{switchback} precisely recovers the growth rate of action in Eq.~\eqref{growthrate} at late time.

This result is identical to the result obtained by Ref.~\cite{Brown_CA&BH} when $\theta=0,z=1$. The calculation in Ref.~\cite{Brown_CA&BH} considers an extra boundary term for the past and future horizon $FOG$ and $EO'H$, which is given by the null limit of the spacelike boundary action
\beq\label{extranullsurface}
I_{\CN}=\lim_{\CS\to\CN} I_{\CS}=\lim_{\rho\to\rho_h}\sgn(\Sigma)\frac{1}{8\pi G}\int_{\Sigma}\sqrt{|h|} K d\Sigma.
\eeq
In the calculation method of this paper, these terms are not present, but a similar contribution is given by considering the effect of the discontinuity of the metric along $U=0$ horizon.

\end{appendix}

\bibliography{emd_complexity_action_bib}

\begingroup\raggedright\begin{thebibliography}{10}

\bibitem{Maldacena_EBH_ADS}
J.~Maldacena, ``Eternal black holes in anti-de Sitter,'' {\em Journal of High
  Energy Physics} {\bf 2003} (2003), no.~04 021,
  \href{http://stacks.iop.org/1126-6708/2003/i=04/a=021}{http://stacks.iop.org/1126-6708/2003/i=04/a=021}.

\bibitem{Ryu_RT}
S.~{Ryu} and T.~{Takayanagi}, ``{Holographic Derivation of Entanglement Entropy
  from the anti de Sitter Space/Conformal Field Theory Correspondence},'' {\em
  Physical Review Letters} {\bf 96} (May, 2006) 181602,
  \href{http://xxx.lanl.gov/abs/hep-th/0603001}{{\tt hep-th/0603001}}.

\bibitem{Raamsdonk_GravityEntanglement}
M.~{van Raamsdonk}, ``{Building up spacetime with quantum entanglement},'' {\em
  General Relativity and Gravitation} {\bf 42} (Oct., 2010) 2323--2329,
  \href{http://xxx.lanl.gov/abs/1005.3035}{{\tt 1005.3035}}.

\bibitem{Swingle_AdS_MERA}
B.~{Swingle}, ``{Entanglement renormalization and holography},'' {\em \prd}
  {\bf 86} (Sept., 2012) 065007, \href{http://xxx.lanl.gov/abs/0905.1317}{{\tt
  0905.1317}}.

\bibitem{Harlow_AdS_QEC}
A.~{Almheiri}, X.~{Dong}, and D.~{Harlow}, ``{Bulk locality and quantum error
  correction in AdS/CFT},'' {\em Journal of High Energy Physics} {\bf 4} (Apr.,
  2015) 163, \href{http://xxx.lanl.gov/abs/1411.7041}{{\tt 1411.7041}}.

\bibitem{Susskind_Complexity}
L.~{Susskind}, ``{Computational Complexity and Black Hole Horizons},'' {\em
  ArXiv e-prints} (Feb., 2014) \href{http://xxx.lanl.gov/abs/1402.5674}{{\tt
  1402.5674}}.

\bibitem{Stanford_ComplexityShocks}
D.~{Stanford} and L.~{Susskind}, ``{Complexity and shock wave geometries},''
  {\em \prd} {\bf 90} (Dec., 2014) 126007,
  \href{http://xxx.lanl.gov/abs/1406.2678}{{\tt 1406.2678}}.

\bibitem{Brown_CA&BH}
A.~R. {Brown}, D.~A. {Roberts}, L.~{Susskind}, B.~{Swingle}, and Y.~{Zhao},
  ``{Complexity, action, and black holes},'' {\em \prd} {\bf 93} (Apr., 2016)
  086006, \href{http://xxx.lanl.gov/abs/1512.04993}{{\tt 1512.04993}}.

\bibitem{Chapman_Complexity_Formation}
S.~{Chapman}, H.~{Marrochio}, and R.~C. {Myers}, ``{Complexity of formation in
  holography},'' {\em Journal of High Energy Physics} {\bf 1} (Jan., 2017) 62,
  \href{http://xxx.lanl.gov/abs/1610.08063}{{\tt 1610.08063}}.

\bibitem{Evenbly_TNGeometry}
G.~{Evenbly} and G.~{Vidal}, ``{Tensor Network States and Geometry},'' {\em
  Journal of Statistical Physics} {\bf 145} (Nov., 2011) 891--918,
  \href{http://xxx.lanl.gov/abs/1106.1082}{{\tt 1106.1082}}.

\bibitem{Pastawski_PerfectTensors}
F.~{Pastawski}, B.~{Yoshida}, D.~{Harlow}, and J.~{Preskill}, ``{Holographic
  quantum error-correcting codes: toy models for the bulk/boundary
  correspondence},'' {\em Journal of High Energy Physics} {\bf 6} (June, 2015)
  149, \href{http://xxx.lanl.gov/abs/1503.06237}{{\tt 1503.06237}}.

\bibitem{Hayden_RandomTensors}
P.~{Hayden}, S.~{Nezami}, X.-L. {Qi}, N.~{Thomas}, M.~{Walter}, and Z.~{Yang},
  ``{Holographic duality from random tensor networks},'' {\em Journal of High
  Energy Physics} {\bf 11} (Nov., 2016) 9,
  \href{http://xxx.lanl.gov/abs/1601.01694}{{\tt 1601.01694}}.

\bibitem{Hartman_BH_interior}
T.~{Hartman} and J.~{Maldacena}, ``{Time evolution of entanglement entropy from
  black hole interiors},'' {\em Journal of High Energy Physics} {\bf 5} (May,
  2013) 14, \href{http://xxx.lanl.gov/abs/1303.1080}{{\tt 1303.1080}}.

\bibitem{Molina-Vilaplana_Hartman-Maldacena}
J.~{Molina-Vilaplana} and J.~{Prior}, ``{Entanglement, tensor networks and
  black hole horizons},'' {\em General Relativity and Gravitation} {\bf 46}
  (Nov., 2014) 1823, \href{http://xxx.lanl.gov/abs/1403.5395}{{\tt 1403.5395}}.

\bibitem{Czech_TNQuotient}
B.~{Czech}, G.~{Evenbly}, L.~{Lamprou}, S.~{McCandlish}, X.-l. {Qi},
  J.~{Sully}, and G.~{Vidal}, ``{Tensor network quotient takes the vacuum to
  the thermal state},'' {\em \prb} {\bf 94} (Aug., 2016) 085101,
  \href{http://xxx.lanl.gov/abs/1510.07637}{{\tt 1510.07637}}.

\bibitem{Nielsen_ComplexityGeometry}
M.~A. {Nielsen}, M.~R. {Dowling}, M.~{Gu}, and A.~C. {Doherty}, ``{Quantum
  Computation as Geometry},'' {\em Science} {\bf 311} (Feb., 2006) 1133--1135,
  \href{http://xxx.lanl.gov/abs/quant-ph/0603161}{{\tt quant-ph/0603161}}.

\bibitem{Jefferson_FreeQFTComplexity}
R.~A. {Jefferson} and R.~C. {Myers}, ``{Circuit complexity in quantum field
  theory},'' {\em Journal of High Energy Physics} {\bf 10} (Oct., 2017) 107,
  \href{http://xxx.lanl.gov/abs/1707.08570}{{\tt 1707.08570}}.

\bibitem{Chapman_FreeQFTComplexity}
S.~{Chapman}, M.~P. {Heller}, H.~{Marrochio}, and F.~{Pastawski}, ``{Towards
  Complexity for Quantum Field Theory States},'' {\em ArXiv e-prints} (July,
  2017) \href{http://xxx.lanl.gov/abs/1707.08582}{{\tt 1707.08582}}.

\bibitem{Brown_CA}
A.~R. Brown, D.~A. Roberts, L.~Susskind, B.~Swingle, and Y.~Zhao, ``Holographic
  Complexity Equals Bulk Action?,'' {\em Phys. Rev. Lett.} {\bf 116} (May,
  2016) 191301, \href{http://xxx.lanl.gov/abs/1509.07876}{{\tt 1509.07876}},
  \href{https://link.aps.org/doi/10.1103/PhysRevLett.116.191301}{https://link.aps.org/doi/10.1103/PhysRevLett.116.191301}.

\bibitem{Lloyd_UltimateBound}
S.~{Lloyd}, ``{Ultimate physical limits to computation},'' {\em \nat} {\bf 406}
  (Aug., 2000) \href{http://xxx.lanl.gov/abs/quant-ph/9908043}{{\tt
  quant-ph/9908043}}.

\bibitem{Carmi_HoloComplexityTimeDep}
D.~{Carmi}, S.~{Chapman}, H.~{Marrochio}, R.~C. {Myers}, and S.~{Sugishita},
  ``{On the Time Dependence of Holographic Complexity},'' {\em ArXiv e-prints}
  (Sept., 2017) \href{http://xxx.lanl.gov/abs/1709.10184}{{\tt 1709.10184}}.

\bibitem{Jordan_NoBound}
S.~P. {Jordan}, ``{Fast quantum computation at arbitrarily low energy},'' {\em
  \pra} {\bf 95} (Mar., 2017) 032305,
  \href{http://xxx.lanl.gov/abs/1701.01175}{{\tt 1701.01175}}.

\bibitem{Couch_NonCommActionGrowth}
J.~{Couch}, S.~{Eccles}, W.~{Fischler}, and M.-L. {Xiao}, ``{Holographic
  complexity and non-commutative gauge theory},'' {\em ArXiv e-prints} (Oct.,
  2017) \href{http://xxx.lanl.gov/abs/1710.07833}{{\tt 1710.07833}}.

\bibitem{Moosa_DivergeRateHoloComplexity}
M.~{Moosa}, ``{Divergences in the rate of complexification},'' {\em ArXiv
  e-prints} (Dec., 2017) \href{http://xxx.lanl.gov/abs/1712.07137}{{\tt
  1712.07137}}.

\bibitem{Carmi_CommentsHC}
D.~{Carmi}, R.~C. {Myers}, and P.~{Rath}, ``{Comments on holographic
  complexity},'' {\em Journal of High Energy Physics} {\bf 3} (Mar., 2017) 118,
  \href{http://xxx.lanl.gov/abs/1612.00433}{{\tt 1612.00433}}.

\bibitem{Couch_NoetherCV}
J.~{Couch}, W.~{Fischler}, and P.~H. {Nguyen}, ``{Noether charge, black hole
  volume, and complexity},'' {\em Journal of High Energy Physics} {\bf 3}
  (Mar., 2017) 119, \href{http://xxx.lanl.gov/abs/1610.02038}{{\tt
  1610.02038}}.

\bibitem{Huang_HoloC&2identities}
H.~{Huang}, X.-H. {Feng}, and H.~{L{\"u}}, ``{Holographic complexity and two
  identities of action growth},'' {\em Physics Letters B} {\bf 769} (June,
  2017) 357--361, \href{http://xxx.lanl.gov/abs/1611.02321}{{\tt 1611.02321}}.

\bibitem{Cai_AdSBH_growthrate}
R.-G. {Cai}, S.-M. {Ruan}, S.-J. {Wang}, R.-Q. {Yang}, and R.-H. {Peng},
  ``{Action growth for AdS black holes},'' {\em Journal of High Energy Physics}
  {\bf 9} (Sept., 2016) 161, \href{http://xxx.lanl.gov/abs/1606.08307}{{\tt
  1606.08307}}.

\bibitem{Cai_ActionChargedBH}
R.-G. {Cai}, M.~{Sasaki}, and S.-J. {Wang}, ``{Action growth of charged black
  holes with a single horizon},'' {\em \prd} {\bf 95} (June, 2017) 124002,
  \href{http://xxx.lanl.gov/abs/1702.06766}{{\tt 1702.06766}}.

\bibitem{Yang_SEC&Cbound}
R.-Q. {Yang}, ``{Strong energy condition and complexity growth bound in
  holography},'' {\em \prd} {\bf 95} (Apr., 2017) 086017,
  \href{http://xxx.lanl.gov/abs/1610.05090}{{\tt 1610.05090}}.

\bibitem{Fu_HCnonlocal}
Z.~{Fu}, A.~{Maloney}, D.~{Marolf}, H.~{Maxfield}, and Z.~{Wang},
  ``{Holographic complexity is nonlocal},'' {\em ArXiv e-prints} (Jan., 2018)
  \href{http://xxx.lanl.gov/abs/1801.01137}{{\tt 1801.01137}}.

\bibitem{Reynolds_CindS}
A.~P. {Reynolds} and S.~F. {Ross}, ``{Complexity in de Sitter space},'' {\em
  Classical and Quantum Gravity} {\bf 34} (Sept., 2017) 175013,
  \href{http://xxx.lanl.gov/abs/1706.03788}{{\tt 1706.03788}}.

\bibitem{Kuang_EMD}
X.-M. {Kuang}, E.~{Papantonopoulos}, J.-P. {Wu}, and Z.~{Zhou}, ``{The Lifshitz
  black branes and DC transport coefficients in massive
  Einstein-Maxwell-dilaton gravity},'' {\em ArXiv e-prints} (Sept., 2017)
  \href{http://xxx.lanl.gov/abs/1709.02976}{{\tt 1709.02976}}.

\bibitem{HIS_complexity&gauge}
K.~Hashimoto, N.~Iizuka, and S.~Sugishita, ``{Time evolution of complexity in
  Abelian gauge theories},'' {\em Phys. Rev.} {\bf D96} (2017), no.~12 126001,
  \href{http://xxx.lanl.gov/abs/1707.03840}{{\tt 1707.03840}}.

\bibitem{Mansoori_CG&BHT}
S.~A.~H. {Mansoori} and M.~M. {Qaemmaqami}, ``{Complexity Growth, Butterfly
  Velocity and Black hole Thermodynamics},'' {\em ArXiv e-prints} (Nov., 2017)
  \href{http://xxx.lanl.gov/abs/1711.09749}{{\tt 1711.09749}}.

\bibitem{Alishahiha_Complexity&FR}
M.~{Alishahiha}, A.~F. {Astaneh}, A.~{Naseh}, and M.~H. {Vahidinia}, ``{On
  complexity for F ( R) and critical gravity},'' {\em Journal of High Energy
  Physics} {\bf 5} (May, 2017) 9,
  \href{http://xxx.lanl.gov/abs/1702.06796}{{\tt 1702.06796}}.

\bibitem{Momeni_fidelity}
D.~{Momeni}, M.~{Faizal}, A.~{Myrzakul}, and R.~{Myrzakulov}, ``{Fidelity
  susceptibility for Lifshitz geometries via Lifshitz Holography},'' {\em ArXiv
  e-prints} (Jan., 2017) \href{http://xxx.lanl.gov/abs/1701.08660}{{\tt
  1701.08660}}.

\bibitem{Huijse_HiddenFS}
L.~{Huijse}, S.~{Sachdev}, and B.~{Swingle}, ``{Hidden Fermi surfaces in
  compressible states of gauge-gravity duality},'' {\em \prb} {\bf 85} (Jan.,
  2012) 035121, \href{http://xxx.lanl.gov/abs/1112.0573}{{\tt 1112.0573}}.

\bibitem{Charmousis_EffLowTempHolo}
C.~{Charmousis}, B.~{Gout{\'e}raux}, B.~{Soo Kim}, E.~{Kiritsis}, and
  R.~{Meyer}, ``{Effective holographic theories for low-temperature condensed
  matter systems},'' {\em Journal of High Energy Physics} {\bf 11} (Nov., 2010)
  151, \href{http://xxx.lanl.gov/abs/1005.4690}{{\tt 1005.4690}}.

\bibitem{Ogawa_HoloFS}
N.~{Ogawa}, T.~{Takayanagi}, and T.~{Ugajin}, ``{Holographic Fermi surfaces and
  entanglement entropy},'' {\em Journal of High Energy Physics} {\bf 1} (Jan.,
  2012) 125, \href{http://xxx.lanl.gov/abs/1111.1023}{{\tt 1111.1023}}.

\bibitem{Dong_AspectsHSV}
X.~{Dong}, S.~{Harrison}, S.~{Kachru}, G.~{Torroba}, and H.~{Wang}, ``{Aspects
  of holography for theories with hyperscaling violation},'' {\em Journal of
  High Energy Physics} {\bf 6} (June, 2012) 41,
  \href{http://xxx.lanl.gov/abs/1201.1905}{{\tt 1201.1905}}.

\bibitem{Evenbly_BranchingMERA_Entanglement}
G.~{Evenbly} and G.~{Vidal}, ``{Scaling of entanglement entropy in the
  (branching) multiscale entanglement renormalization ansatz},'' {\em \prb}
  {\bf 89} (June, 2014) 235113, \href{http://xxx.lanl.gov/abs/1310.8372}{{\tt
  1310.8372}}.

\bibitem{Swingle_AreaLawEntanglementThermo}
B.~{Swingle} and J.~{McGreevy}, ``{Area law for gapless states from local
  entanglement thermodynamics},'' {\em \prb} {\bf 93} (May, 2016) 205120,
  \href{http://xxx.lanl.gov/abs/1505.07106}{{\tt 1505.07106}}.

\bibitem{Lehner_Grav_Nullboundary}
L.~{Lehner}, R.~C. {Myers}, E.~{Poisson}, and R.~D. {Sorkin}, ``{Gravitational
  action with null boundaries},'' {\em \prd} {\bf 94} (Oct., 2016) 084046,
  \href{http://xxx.lanl.gov/abs/1609.00207}{{\tt 1609.00207}}.

\bibitem{DHoker_MagBraneBH}
E.~{D'Hoker} and P.~{Kraus}, ``{Magnetic brane solutions in AdS},'' {\em
  Journal of High Energy Physics} {\bf 10} (Oct., 2009) 088,
  \href{http://xxx.lanl.gov/abs/0908.3875}{{\tt 0908.3875}}.

\bibitem{Roberts_ScramblingHSV}
D.~A. {Roberts} and B.~{Swingle}, ``{Lieb-Robinson and the butterfly effect},''
  {\em ArXiv e-prints} (Mar., 2016)
  \href{http://xxx.lanl.gov/abs/1603.09298}{{\tt 1603.09298}}.

\bibitem{Maldacena_ChaosBound}
J.~{Maldacena}, S.~H. {Shenker}, and D.~{Stanford}, ``{A bound on chaos},''
  {\em Journal of High Energy Physics} {\bf 8} (Aug., 2016) 106,
  \href{http://xxx.lanl.gov/abs/1503.01409}{{\tt 1503.01409}}.

\bibitem{Haegeman_WaveletRigor}
J.~{Haegeman}, B.~{Swingle}, M.~{Walter}, J.~{Cotler}, G.~{Evenbly}, and V.~B.
  {Scholz}, ``{Rigorous free fermion entanglement renormalization from wavelet
  theory},'' {\em ArXiv e-prints} (July, 2017)
  \href{http://xxx.lanl.gov/abs/1707.06243}{{\tt 1707.06243}}.

\end{thebibliography}\endgroup
\bibliographystyle{ucsd}
\end{document}